\documentstyle[epsfig,aps]{revtex}

\newenvironment{figures}[1]%
{\begin{list}{}{\settowidth{\labelwidth}{#1}
  \setlength{\leftmargin}{\labelwidth}
  \addtolength{\leftmargin}{\labelsep}
  \setlength{\parsep}{1ex plus0.7ex minus0.7ex}
  \setlength{\itemsep}{0.8ex}
  }}{\end{list}}

\newcommand{\half}{\mbox{\scriptsize{$\frac{1}{2}$}}}

\newcommand{\be}{\begin{equation}}
\newcommand{\bel}[1]{\begin{equation}\label{#1}}
\newcommand{\bela}[1]{\begin{eqnarray}\label{#1}}
\newcommand{\bea}{\begin{eqnarray}}
\newcommand{\ee}{\end{equation}}
\newcommand{\eela}{\end{eqnarray}}
\newcommand{\eea}{\end{eqnarray}}

\begin{document}

\title {
{\small
\hspace{11cm} NIKHEF 97--020 \\
\hspace{11cm} hep-ph/9707262 }\\ \vspace{1cm}
{\bf The BFKL Pomeron with running coupling constant:\\
how much of its hard nature survives?}}

\author{
L.P.A. Haakman
$^{a}$,
O.V. Kancheli
$^{b}$,
J.H. Koch
$^{a,c}$}

\address{
$^{a}$
National Institute for Nuclear Physics
and High Energy Physics (NIKHEF), \\
P.O. Box 41882, NL-1009 DB Amsterdam, The Netherlands \\
$^{b}$
Institute of Theoretical and Experimental Physics, \\
B. Cheremushinskaya 25, 117 259 Moscow, Russia\\
$^{c}$
Institute for Theoretical Physics, University of Amsterdam}

\date{7 July 1997}

\maketitle
\begin{abstract}
  We discuss the BFKL equation with a running gauge coupling 
and identify in its solutions the contributions
originating from different transverse momentum scales.
We show that for a running coupling constant the distribution
of the gluons making up the BFKL Pomeron shifts to smaller 
transverse momenta so that the dominant part of Pomeron 
can have a nonperturbative origin. 
It is demonstrated how this soft physics enters into the BFKL 
solution through the boundary condition.
We consider two kinematical regimes leading to different behaviour 
of the rapidity and transverse momentum dependence of the 
gluon distribution.
In the diffusion approximation to the BFKL kernel with running 
$\alpha_s$, we find a sequence of poles which replaces the cut 
for fixed $\alpha_s$.
The second regime corresponds to the singular part of the kernel, 
which gives the dominant contribution in the limit of very large 
transverse momenta. 
Finally, a simple more general picture is obtained for the 
QCD Pomeron in hard processes: it is of soft, nonperturbative
nature, but has hard ends of DGLAP-type.
\end{abstract}


\section*{\bf 1. Introduction}
\renewcommand{\theequation}
           {1.{\arabic{equation}}}\setcounter{equation}{0}
The description of high energy reactions in terms of the exchange of
a Pomeron and of secondary Reggeons continues to be very successful.
Ever since the arrival of QCD, the challenge has of course been to 
provide an elementary, microscopic description of the Regge theory.
Nonperturbative aspects have so far made it impossible to
have a complete QCD description. 
However, in the extreme limit of very high rapidities and large 
transverse momenta, it was possible to model the Pomeron perturbatively
by the exchange of a gluon ladder and obtain 
an analytic expression under certain kinematical conditions. 
This was called the BFKL Pomeron \cite{BFKL}, which was considered 
to be of somewhat academic interest. 
The recent HERA experiments, which showed a very rapid 
increase of the structure function $F_2(x,Q^2)$ in the limit of
small $x$, have renewed the interest in the BFKL Pomeron, since it  
predicts such a behaviour. 
This inspired the hope that this and other features resulting from the
BFKL Pomeron already start to appear at existing energies.

It is commonly assumed that the Pomeron in QCD is dominated by 
gluons. 
Due to its quantum numbers it has to consist of at least two gluons, 
which would result in a constant cross section as a function of energy
\cite{2g_pom}. 
The admixture of multigluon states and other, nonperturbative 
effects then should lead to the Pomeron, with glueball states 
on its trajectory. 
This trajectory should have the observed rather small slope 
$\alpha_P'$ and ``supercritical'' intercept $\alpha_P(0)>1$.
The calculations done so far, while not complete, were at least 
able to confirm this concept of the Pomeron.
 
Due to the smallness of the gauge coupling, the two gluon state is 
expected to dominate the Pomeron in reactions involving high 
transverse momenta. 
However, at high energies or correspondingly for high rapidities,
a simple perturbative calculation of the two gluon exchange is not 
sufficient and interactions between the two gluons must be taken into
account. 
This is done in the well-known BFKL Pomeron $\cite{BFKL}$ in leading 
logarithmic approximation (LLA) in $\log (1/x)$ with a fixed QCD 
coupling constant. 
This ``hard'' Pomeron can be represented by the exchanges of gluon 
ladders. 
 
An interesting and characteristic feature of the BFKL Pomeron is 
the diffusion in the transverse momenta (in fact in the logarithms) 
of the gluons as the rapidity increases along the gluonic ladder. 
It can lead to a growth of the mean transverse momentum with energy
and one expects that at very high energies the ``hard  Pomeron'' 
dominates in processes with high transverse masses.
In this region of high virtualities also the DGLAP evolution might be 
applied.
For the diffusion property at fixed $\alpha_s$ it is essential  
that the BFKL equation is scale invariant.
Clearly it will be interesting to examine how the situation will change
after this invariance is broken.
Since the main source of the scale invariance violation is 
the dependence of the effective gauge coupling on the QCD scale
$\Lambda$, the most obvious way to study this question is to include 
the running gauge coupling in the BFKL equation. 
Evidently there are other higher order corrections: one part
concerns the BFKL gluon ladder itself \cite{next_alpha}, but one 
expects that they should not change its qualitative features 
much more; other corrections involve multiple Reggeon diagrams 
\cite{eff_act}.
Some aspects of these questions have already been 
considered in Refs.\cite{Lip2,Kwe1,HanRoss,NiZa1,Levin2,BrPa,CimCa}.

It is the aim of this paper to analyse the generalization of the BFKL 
approach to a running coupling constant in the most straightforward 
fashion and to arrive at a comprehensive physical interpretation.
Our numerical study in Ref.\cite{jlo} of the BFKL equation with a
running coupling constant has already shown that the BFKL Pomeron is
of purely perturbative character only at the ends of the gluon ladder, 
while becoming progressively soft towards the middle. 
We study here the presence of a running coupling constant in more 
detail in an analytical fashion and, in addition, examine the 
singularity structure of the Pomeron in the complex angular momentum 
plane.

The outline of our paper is as follows.
We begin with a review of the original BFKL equation with fixed coupling
constant in Section 2.
Then in Section 3 we present the exact solution of the BFKL equation
with running $\alpha_s$, chosen in the simplest form for the full range 
of transverse momenta and without an infrared cut-off.
The main result is that we find an essential singularity in the 
complex angular momentum plane, independent of any transverse 
momentum scale, that leads to an energy dependence for the cross 
sections of the form ${\rm e}^{c{\sqrt y}}$.  
This result is further explored in the following sections.
In Section 4 we examine a diffusion-like approximation to the full
operator equation. 
We find that this diffusion solution contains an infinite sequence of 
poles in the complex variable $\omega$, the conjugate of the rapidity $y$.
This resembles the series of poles found earlier by Lipatov 
$\cite{Lip2}$, which replaces the cut in the case of fixed $\alpha_s$. 
We examine the residues of these poles and their importance in hard 
processes. 
A simple quantum-mechanical analogue is presented to make the 
essential characteristics of the diffusion solution more 
transparent.
In Section 5, the contribution from the singular part of the 
BFKL kernel is discussed. 
It represents large increments in the transverse momenta of 
the gluons along the ladder and is shown to lead to a 
DGLAP-like behaviour.
Based on the features we found for the BFKL Pomeron with running 
$\alpha_s$, we arrive at a simple more general model for the Pomeron;
it reflects all the essential properties of the exact solution 
of Section 3.
Its main point is that the Pomeron is of soft, nonperturbative
origin, but can have hard ends, when probed by a hard device 
as in deep-inelastic scattering.
In the limit of very large rapidities the hard ends become
small compared to the soft part.
Applications to onium-onium scattering are considered.
A summary and our conclusions are contained in Section 6.
The main conclusions of this paper were already reported  in 
Ref.\cite{erice}.


\section*{\bf 2. BFKL equation with fixed coupling constant}
\renewcommand{\theequation}
           {2.{\arabic{equation}}}\setcounter{equation}{0}
In this section we review shortly the BFKL equation with fixed
coupling constant $\alpha_s$. 
It has been experimentally established that the hard cross sections
corresponding to the exchange of vacuum quantum numbers have a 
power-law increase as function of the CMS energy $\sqrt{s}$ according 
to $\sigma \sim s^{\Delta_{exp}}$ with $\Delta_{exp}\sim 0.3$.
A simple two-gluon exchange as model for the hard Pomeron \cite{2g_pom}
however leads to constant total cross sections, and diagrams leading to
large $\log (s)$ contributions must be taken into account. 
BFKL included such more complex diagrams which could be resummed and 
represented as exchanges of effective gluon ladders. 
In these ladders non-local gauge invariant Lipatov vertices and 
reggeized gluon propagators are the building blocks (see Fig.1). 
This was done under the assumption of ``multiregge kinematics'', 
where the rapidity monotonically increases along the ladder in 
large steps, $\bar{{\delta} } \gg 1$.
These ladder-like configurations make the cross sections increase
fast with energy.
BFKL performed their calculations for fixed coupling constant
$\alpha_s$ in leading logarithmic approximation, {\em i.e.}
\bel{LLA}
\alpha_s \ll 1 ~~;~~~ \alpha_s y \sim 1~~, 
\ee
where the rapidity interval between the ends of the ladder, $y$, is 
related to the total energy, according to $y \sim \log(s)$.
The simple ladder structure of the diagrams makes it possible
to write a Bethe-Salpeter type of equation for this amplitude.
This leads to a linear integro-differential equation, known as the 
BFKL equation for the low $x$ behaviour of the unintegrated
gluon distribution in the nucleon:
\bel{BFKLeq}
\frac{\partial f(y,k_{\perp}^2)}{\partial y} ~=~
\frac{3 \alpha_s}{\pi} \int_{0}^\infty \frac{d {q}_{\perp}^2}
{q_{\perp}^2} k_{\perp}^2
\left[\frac{f(y,q_{\perp}^2)-f(y,k_{\perp}^2)}{|q_{\perp}^2 - k_{\perp}^2|} + \frac{f(y,k_{\perp}^2)}
{\sqrt{4q_{\perp}^4+k_{\perp}^4}}\right]\equiv \bar{\alpha}_s\,
{\cal L}(k_{\perp}^2)\otimes f(y,k_{\perp}^2)~~, 
\ee
where we defined the rapidity as $y=\log(x_0/x)$ with $x_0$ a higher $x$
value at which the initial condition should be given, 
and $\bar{\alpha}_s= {3\alpha_s}/{\pi}$.
In Eq.(\ref{BFKLeq}) the integration over the angle between $k_{\perp}$
and $q_{\perp}$ has already been performed. 
The function $f(y,k_{\perp}^2)$ is proportional to the 
imaginary part of the virtual gluon forward scattering amplitude 
with vacuum quantum numbers in the $t$-channel; 
the first term of the integrand corresponds to the emission of real 
gluons while the second term originates from the virtual part of the 
amplitude.
In a reaction with characteristic scale $Q^2$, one probes the 
integrated gluon distribution,
\bel{gluon}
G(y,Q^2)=\int^{Q^2}_{\mu^2} \frac{dk_{\perp}^2}{k_{\perp}^2}\,
f(y,k_{\perp}^2)~~.
\ee

Since the BFKL equation is scale invariant, it can be
solved in a simple way using Mellin and Laplace transformations 
with respect to $k_{\perp}^2$ and $y$.
Therefore we define
\bea
{\tilde{f}}(y,\beta)&=&\int_0^{\infty}\frac{dk_{\perp}^2}{k_{\perp}^2}
\left(\frac{k_{\perp}^{2}}{m^2}\right)^{-\beta}f(y,k_{\perp}^2)
=\int_{-\infty}^{\infty} du~ {\rm e}^{-\beta u}f(y,u)~~, 
\label{Mellindef}\\ 
\bar{f}(\omega,u)&=&\int_{0}^{\infty}{dy}~
{\rm e}^{-\omega y }f(y,u)~~,\label{Laplacedef1} \\
F(\omega,\beta)&=&\int_0^{\infty}{dy}~ {\rm e}^{-\omega
y}{\tilde{f}}(y,\beta)~~,\label{Laplacedef2}
\eea
where $u=\log({k}_{\perp}^2/m^2)$, a definition we use throughout
the paper, and where $m$ is a scale introduced in order to make the 
transformed distribution $\tilde{f}(y,\beta)$ in Eq.(\ref{Mellindef}) dimensionless.
The variable $\omega$, conjugate to the total rapidity interval 
$y$, has a physical interpretation relating it to the (complex) 
angular momentum: $\omega = j - 1$.
Applying both transformations to Eq.(\ref{BFKLeq}) leads to the algebraic 
equation
\bel{algb1}
\omega F(\omega,\beta)=\tilde{f}(0,\beta) +
\bar{\alpha}_s \tilde{\cal L}(\beta) F(\omega,\beta)~~,
\ee
with the solution
\bel{solBF} F(\omega,\beta)=\frac{\tilde{f}(0,\beta)}{\omega
- \bar{\alpha}_s \tilde{\cal L}(\beta)}~~,
\ee
where
\bel{kernel}
\tilde{\cal L}(\beta)=2\Psi(1)-\Psi(\beta)-\Psi(1-\beta)~~,~~~~
 \Psi(\beta) = \frac{d \log\Gamma(\beta)}{d \beta}~~,
\ee
is the eigenvalue spectrum of the BFKL integral operator in 
Eq.(\ref{BFKLeq}).
The Mellin transformation with respect to ${k}_{\perp}$ was only 
well defined for $0<{\rm Re}\,\beta<1$.
In this range of $\beta$ the function $\tilde{\cal L}$ has a 
minimum at $\beta =1/2$ while it is singular at the edges $\beta=0$ 
and $\beta=1$ (see Fig.2).
The limit $\beta\rightarrow 0$ corresponds to the hard limit, where
$u$ is large.   

We take the inverse transformations over $\beta$ and 
$\omega$ of Eq.(\ref{solBF}) to get back to $f(y,u)$
\bela{Mellininv}
{\tilde{f}}(y,\beta)&=&\frac{1}{2\pi i}
\int_{c-i\infty}^{c+i\infty}{d\omega}~ {\rm e}^{\omega
y}F(\omega,\beta)~~,\\
f(y,u)&=&\frac{1}{2\pi i}\int_{d-i\infty}^{d+i\infty}
 d \beta~ {\rm e}^{u\beta} {\tilde{f}}(y,\beta)~~, 
\eela
with $c$ a real number to the right of all singularities in
$\omega$ and $d$ an appropriately chosen real number between $0$
and $1$.
This yields the solution of the BFKL equation
\bel{sol1}
f(y,u)=\int_{-\infty}^{\infty}du' ~f(0,u')\, {\cal{G}}_f(y,u-u')~~,
\ee
where $f(0,u')$ is the boundary condition at $y=0$, and
with the Green's function
\bel{Greenfix}
{\cal{G}}_f(y,u)=\frac{1}{2\pi i}
\int_{\half-i\infty}^{\half+i\infty}d \beta~ 
{\exp\left(\beta u+y\bar{\alpha}_s\tilde{\cal{L}}(\beta)\right)}~~,
\ee
where we took $d = 1/2$ due to the pinch of the 
$\beta$ contour at the point $\beta=1/2$. 
This pinch results in a leading singularity in Eq.(\ref{solBF}) at 
$\omega=\bar{\alpha}_s \tilde{{\cal{L}} }(1/2)$.
For large $y$ the saddle point method can be used to calculate 
explicitly the Green's function.
With the help of the expansion of the kernel around $\beta=1/2$,
\bel{kern.exp}
\bar{\alpha}_s\tilde{\cal{L}}(\beta)\approx\Delta + 
B (\beta-\frac{1}{2})^2~~~;~~~
\Delta=4\bar{\alpha}_s\log(2) ~~,~~B=14\bar{\alpha}_s\zeta(3)~~,
\ee
one then obtains
\bela{sol2}
{\cal{G}}_f(y,u) &\approx&\frac{1}{\sqrt{4\pi B y e^{-u}}}
\exp\left(y\Delta -\frac{u^2}{4 B y}\right)~~.
\eela
The character of the corresponding singularity in $\omega$ 
and its dependence on $u$ can be more easily seen in 
the $(\omega,u)$-representation where the Green's function
takes the form
\bel{sol3}
{\cal{G}}_f({\omega},u)\approx\frac{1}{\sqrt{4B e^{-u}(\omega -\Delta)}} \,
\exp{\left[-u \sqrt{\frac{\omega-\Delta}{B}} \right]}~~,
\ee
which shows the branch point singularity at $\omega = \Delta$.

From Eqs.(\ref{sol1}) and (\ref{sol2}) we see the two characteristic 
features of the BFKL Pomeron.
First, the Pomeron intercept $\Delta\simeq 2.6\,\alpha_s$ is relatively
large, of order $0.5$ for a realistic $\alpha_s \sim 0.2$.
Therefore the observed energy dependence of cross sections by the
HERA experiments, $\sigma\sim s^{{\Delta}_{exp}}$, was seen as a possible 
sign of the appearance of the BFKL Pomeron in hard scatterings.
Second, the diffusion of the  $u$ distribution as $y$ grows:
$<(u-\bar{u})^2> \,\sim \,4By$.  
As will be seen later, this important diffusion property is a 
consequence of scale invariance when $\alpha_s$ is fixed.

We conclude this section with two comments about the 
applicability of the BFKL Pomeron in LLA. 
Due to the diffusion of the $u$ distribution, it can happen that as 
$y\rightarrow \infty $ one reaches regions of small 
transverse momenta, where non-perturbative effects take place. 
For the perturbative BFKL Pomeron to be applicable, 
these contributions should be small.
One can simply exclude transverse momenta below some fixed 
$k_{\perp}^{min}$ in all gluon ladder diagrams by introducing a 
cut-off in the integral that enters into the BFKL equation 
\cite{ColLan1,MFR}.
Another way to achieve this is to consider the BFKL Pomeron 
in onium-onium scattering.
In this case there is a purely perturbative regime near both ends of
the gluon chain due to the small transverse sizes of the onia,
$ R_{\perp}^2\sim 1/k_{\perp}^2$.
They can be chosen sufficiently small, so that the transverse 
momenta in the gluon ladder do not reach the infrared region in 
the entire rapidity interval between the two onia.
In the numerical examples in Ref.\cite{jlo},
the influence of the small $k^2_{\perp}$ region on the gluon ladder 
was found to be not large for fixed $\alpha_s$; the gluons in the ladder 
are in the mean far from the soft region.
However, for running $\alpha_s$ the diffusion to the infrared region 
was shown to proceed much faster. 

Concerning the assumption of multiregge kinematics, we can
make the following estimate for the BFKL Pomeron, based on a
general property, independent from the underlying QCD dynamics.
For a ladder diagram with vector particle exchange the cross 
section grows as ${\rm e}^{y\Delta}$.  
Expanding this exponential function as $\sum_n (y \Delta)^n/n!$, one
obtains the cross section in terms of contributions from ladders 
with $n$ rungs and a corresponding mean multiplicity 
$\bar{n} = y \Delta$.  
As result, the mean rapidity interval between neighbouring rungs 
is $\bar{\delta} \equiv <y_i- y_{i+1}> \simeq \Delta^{-1}$.
This quantity is not large: for $\alpha_s\simeq 0.2$ one has
$\bar{\delta} \sim 2$ or $\Delta \simeq 0.5$.  
However, the BFKL was derived under the assumption of multiregge kinematics,  $\bar{\delta} \gg 1$.
If we consider ladders making up the Pomeron with a lower cut-off, 
$\delta_0 \simeq 1$, for the rapidity difference $\delta = y_i- y_{i+1}$, 
a sizeable fraction of the gluons are eliminated.
This is due to the fact that the gluons are concentrated at small 
rapidity differences as was confirmed numerically in Ref.\cite{jlo}.  
For $\alpha_s = 0.2$ it was found that the new $\Delta$
characterizing the resulting energy dependence of the cross section 
becomes smaller by approximately a factor of two.  

\section*{\bf 3. Exact solutions of BFKL with running $\alpha_s$}
\renewcommand{\theequation}
           {3.{\arabic{equation}}}\setcounter{equation}{0}
In this section we include a running coupling constant in
the BFKL equation. 
The procedure to do this is not unique.
It is often required that the so-called ``bootstrap condition" 
\cite{Lip2} is obeyed. 
This consistency condition follows from the requirement that the 
solution of the BFKL equation in the gluon colour octet channel 
should be that of the reggeized gluon itself.
Further one should check that in the limit of very high transverse 
momenta the double logarithmic approximation of the DGLAP equation
is obtained.
Nevertheless, many ways to introduce a running gauge coupling 
$\alpha_s$ in the BFKL equation remain possible, but they are expected 
only to result in non-essential corrections.

We include a running coupling constant in the BFKL 
equation by using instead of the fixed one in Eq.(\ref{BFKLeq})
the perturbative expression
\bel{runningalpha}
\alpha_s(k_{\perp}^2) =\frac{1}{b\log(k_{\perp}^2/\Lambda^2)} ~~,
\ee
where $\Lambda$ is the QCD scale parameter and $b=(33-2N_f)/12\pi$.
This choice was also made in Ref.\cite{Lip2,Levin2}. 
We put the running coupling constant under the integral in the
following way:
\bel{BFKLeqrun1}
\frac{\partial f(y,k_{\perp}^2)}{\partial y} ~=~
\frac{3}{\pi} \int_{0}^\infty \frac{d {q}_{\perp}^2}
{q_{\perp}^2} k_{\perp}^2
\left[\frac{\alpha_s(q_{\perp}^2)\,f(y,q_{\perp}^2)-
\alpha_s(k_{\perp}^2)\,f(y,k_{\perp}^2)}{|q_{\perp}^2 - k_{\perp}^2|} + \frac{\alpha_s(k_{\perp}^2)f(y,k_{\perp}^2)}
{\sqrt{4q_{\perp}^4+k_{\perp}^4}}\right)~~.
\ee
This has precisely the structure of the original BFKL equation 
when we write $h(y,k_{\perp}^2) =\alpha_s(k_{\perp}^2)
f(y,k_{\perp}^2)$:
\bel{BFKLeqrun2}
\frac{\partial h(y,k_{\perp}^2)}{\partial y} ~=~
\alpha_s(k_{\perp}^2) \,
{\cal L}(k_{\perp}^2)\otimes h(y,k_{\perp}^2)~~.
\ee
It is also possible to leave the running coupling
constant $\alpha_s(k_{\perp}^2)$ outside the integral. 
In this case we obtain a modified BFKL equation as in 
Eq.(\ref{BFKLeqrun2}), but now for $f(y,k_{\perp}^2)$ instead 
of $h(y,k_{\perp}^2)$.
Therefore it yields a gluon distribution with an additional 
factor $\alpha_s(k_{\perp}^2)$ as compared to our choice 
where it is put under the integral. 
As we will see later, our method is the one that leads to the correct
behaviour for large transverse momenta. 

We begin with a qualitative argument in order to show how
including a running $\alpha_s$ in the BFKL equation can modify
the characteristic BFKL features.
For constant $\alpha_s$ the large rapidity behaviour 
of the total cross section predicted by the BFKL equation is given 
by $\sigma(y) \sim \exp(y \alpha_s 12 \log(2)/\pi)$.
The $\alpha_s$ entering this expression should be taken at some 
fixed, effective value of $k_{\perp}$ that must be of the order of 
the mean $k_{\perp}$ in the gluonic ladder representing the Pomeron.
However, for moderate transverse momenta at the lower end of the gluon ladder
and a high virtuality at the top of the ladder ({\em e.g.} in deep 
inelastic scattering), the mean $\log(k_{\perp}^2)$ changes along 
the ladder and is proportional to $\sqrt{y}$. 
Therefore, a simple estimate for the effect of a running 
coupling constant can be obtained by using
$\alpha_s \sim 1/\sqrt{y}$, leading to
a cross section  $\sigma(y) \sim {\rm e}^{c\sqrt{y}}$.
This behaviour is also what we find in the 
double logarithmic approximation of the DGLAP equation.
As we will see later, in the complex angular momentum 
plane $\omega=j-1$ it corresponds to an essential singularity of
$\exp{(1/\omega})$ type.  

By using $\alpha_s$, Eq.(\ref{runningalpha}), for all $k_{\perp}$
we partly suppress the infrared region, which is in line with our goal 
to investigate the influence of a running $\alpha_s$ on the 
structure of the BFKL Pomeron at large $k_{\perp}$.
Qualitatively this may be understood from the fact that the contribution 
of a ladder for vector exchange, with $n$ rungs and in a limited region of
integration over $k_{\perp}$, results in amplitudes of the form
\bel{n_term}
\left[\alpha_s(\bar{\mu}^2)\,{\cal D}\right]^n 
\frac{y^n}{n!}~~,~{\rm with}~~~
\alpha_s(\bar{\mu}^2)\,{\cal D}(\mu_1, \mu_2)
=\int_{k_{\perp}^2 >\mu_1^2}^{k_{\perp}^2 < \mu_2^2}
 d^2 k_{\perp}\, 
\alpha_s(k_{\perp}^2)\left[...\right]~~~~~~(~{\cal D} > 0~)~~,
\ee
where $\bar{\mu}$ the mean scale at which the running
coupling constant has to be taken.
In this way one obtains a Pomeron intercept 
with $\Delta_0\sim{\alpha_s}(\bar{\mu}^2){\cal D}$.
Hence we see that the simplest way to suppress
the soft contribution is to make $\alpha_s$ negative for
small $k_{\perp}$ so that the corresponding singularity moves to
the left half $\omega$-plane.
Therefore it is not bad that the standard expression for 
$\alpha_s$, Eq.(\ref{runningalpha}), has this property
for $k_{\perp}^2<\Lambda^2$.
Note also that the singularity in Eq.(\ref{runningalpha}) is 
integrable at $k_{\perp}^2=\Lambda^2$.

As in the case of constant $\alpha_s$, we solve the BFKL equation with 
running $\alpha_s$, Eq.(\ref{BFKLeqrun2}), by using the Laplace and Mellin transformations of Eqs.(\ref{Mellindef}) and (\ref{Laplacedef2}).
In the Mellin tranformation of Eq.(\ref{Mellindef}) 
we set $m^2=\Lambda^2$.
With the help of the simple identity
\bel{trivial}
\left(\frac{k_{\perp}^{2}}{\Lambda^2}\right)^{-\beta}
\log\left(\frac{k_{\perp}^2}{\Lambda^2}\right)=
-\frac{\partial}{\partial \beta}
\left(\frac{k_{\perp}^{2}}{\Lambda^2}\right)^{-\beta}~~,
\ee
we then obtain for the Mellin transformed differential equation 
for the distribution $h(y,u)$
\bel{BFKLtr1}
-\frac{\partial^2 {\tilde{h}}(y,\beta)}{\partial y \partial
\beta}=\tilde{\alpha}_s
\tilde{\cal L}(\beta){\tilde{h}}(y,\beta)~~,
\ee
where $\tilde{\cal L}(\beta)$ is the standard BFKL kernel, 
Eq.(\ref{kernel}), and we denoted $\tilde{\alpha}_s=3/\pi b$.
Further we perform the Laplace transform over the rapidity $y$ 
which yields for Eq.(\ref{BFKLtr1})
\bel{BFKLtr2}
\frac{\partial{{H}}(\omega,\beta)}{\partial
\beta}=\frac{1}{\omega}\left\{H_0(\beta)-\tilde{\alpha}_s
\tilde{\cal L}(\beta)H(\omega,\beta)\right\}~~,
\ee
with $H_0(\beta)$ a function determined from the initial condition
at $y=0$,
\bel{F0}
H_0(\beta)=\frac{\partial \tilde{h}(0,\beta)}{\partial\beta}~~.
\ee
The homogeneous solution of Eq.(\ref{BFKLtr2}) is
\bel{HomSol}
{H}_{H}(\omega,\beta)={H}_H(\omega,\beta_0)
\,\exp{\left[-\frac{\tilde{\alpha}_s}{\omega}\left({\cal R(\beta)}-
{\cal R}(\beta_0)\right)\right]}~~,
\ee
where
\bel{funcR}
{\cal R(\beta})=\int^{\beta} d\beta'
\tilde{\cal L}(\beta')=2\Psi(1)\beta ~-~ \log
\left[\frac{\Gamma(\beta)}{\Gamma(1-\beta)}\right]~~.
\ee
The function ${\cal R(\beta})$ is plotted in Fig.2 .
Thus the total solution becomes
\bel{InhSol}
H(\omega,\beta)={H}_H(\omega,\beta)+\frac{1}{\omega}
\int_{\beta_0}^{\beta}d\beta'\, H_0(\beta')\,
\frac{{H}_{H}(\omega,\beta)}{{H}_{H}(\omega,\beta')}~~.
\ee 
This is the exact solution of the BFKL equation with a running 
coupling constant.

It is essential that Eq.(\ref{InhSol}) has only a fixed 
singularity at $\omega=0$, {\em i.e.} not depending on $\beta$.
This is in contrast to the situation with a fixed $\alpha_s$ 
where the analogous solution in the $(\omega,\beta )$-representation 
is given in Eq.(\ref{solBF}), and results in a $\beta$ dependent 
$\omega$-pole.
For running $\alpha_s$ there are no pinch and endpoint singularities 
when carrying out the inverse tranformations needed to obtain 
the solution as function of $y$ and $u$, since also the 
singularities in $\beta$ are not dependent on $\omega$. 

For the purpose of studying the $y$ dependence or, equivalently, 
the $\omega$ structure, we want to focus on the first term in 
Eq.(\ref{InhSol}).
The second term does not introduce any new singularities in
$\omega$ since its integrand is proportional to the first term and
singularities in $\omega$ are independent of $\beta$.
The solution of the BFKL equation with running coupling constant
then becomes in the $(\omega,u)$-representation
\bel{sol6}
\bar{h}(\omega,u)=\bar{h}(\omega,u_0)\,
\frac{\bar{K}_r(\omega,u)}{\bar{K}_r(\omega,u_0)}
~~,
\ee
with $\bar{K}_r(\omega,u)$ given by 
\bel{Melliny}
\bar{K}_r(\omega,u)=\frac{1}{2\pi i}
\int_{d-i\infty}^{d+i\infty}
d\beta\,\exp{\left[\beta u-\frac{\tilde{\alpha}_s}
{\omega}{\cal R(\beta)}\right]}~~.
\ee
The function $\bar{h}(\omega,u_0)$ is determined by 
the boundary condition for $u$ at $u_0$.
The function $h(y,u)$, which will yield the gluon distribution 
as function of rapidity and transverse momentum, can finally be 
obtained from the inverse Mellin transformation of 
Eq.(\ref{sol6}) with respect to $\omega$. 

We now discuss an explicit solution of the BFKL equation with 
running coupling constant at very hard scales, appropriate for 
deep inelastic scattering. 
In this case we expect to find the double logarithmic approximation 
in $\log(1/x)$ and $\log(k_\perp^2)$ of DGLAP. 
The large transverse momentum behaviour is determined by the 
low $\beta$ limit of Eq.(\ref{Melliny}).
We can then use  
\bel{approx}
\left[\,\Gamma(\beta)\,\right]^{\tilde{\alpha}_s/\omega}
\simeq \beta^{- \tilde{\alpha}_s/\omega}~~,
\ee
and the inverse Mellin transformation in Eq.(\ref{Melliny})
can be easily evaluated.  
Only the singularity at $\beta\sim 0$ is important;
contributions to $\bar{h}(\omega,u)$ from cuts starting at
$\beta=-1,-2, \ldots$ have additional factors ${\rm e}^{-u}$, 
${\rm e}^{-2u}, \ldots$ and thus can be neglected at 
large $u$.
After deformation of the path along the imaginary axis
we can use Hankel's contour integral formula
for $\Gamma$-functions \cite{AbraSte} and obtain
\bel{APsol1}
\bar{h}(\omega,u)\simeq 
\frac{\bar{h}(\omega,u_0)}{\bar{K}_r(\omega,u_0)}
\frac{1}{2\pi i} \int_{{\cal C}}{d\beta}~
{\rm e}^{u\beta+2\Psi(1)\beta\tilde{\alpha}_s/\omega}
\left(\frac{1}{\beta}\right)^{ \tilde{\alpha}_s/\omega} 
\simeq\,
\frac{\bar{h}(\omega,u_0)}{\bar{K}_r(\omega,u_0)}\frac{1}
{\Gamma(\tilde{\alpha}_s/\omega)}\,u^{-1+\tilde{\alpha}_s/\omega}~~.
\ee
The interpretation of this solution is more transparent if we
rewrite it in the more familiar form
\bel{AnoDim}
\bar{f}(\omega,u)=\frac{1}{\alpha_s(u)}\bar{h}(\omega,u)\sim
\bar{f}(\omega,u_0)\left[\frac{1}
{\alpha_s(u)}\right]^{\frac{1}{2\pi b}\gamma_{gg}(\omega)}~~~~~~
{\rm with}~~~~\gamma_{gg}(\omega)=\frac{6}{\omega}~~.
\ee
In this form for $\bar{f}(\omega,u)$, we recognize the anomalous 
dimension $\gamma_{gg}(\omega)$, thus showing that we indeed
arrive in this limit at the DGLAP solution.
As was remarked earlier, we could have placed the running
coupling constant outside the integral in the BFKL equation.
We would then have obtained as a result of that equation directly
the function $\bar{f}(\omega, u)$, the gluon distribution. 
In that case the gluon distribution differs by an extra factor 
$\alpha_s(u)\sim 1/u$, yielding a discrepancy from DGLAP. 

An important ingredient in the solutions is the boundary 
condition at low transverse momenta $\bar{f}(\omega,u_0)$ 
which provides the function $\bar{h}(\omega,u_0)$ in Eq.(\ref{sol6}).
Poles in $\bar{f}(\omega,u_0)$ at $\omega = \Delta_i$ result in a 
rapidity dependence ${\rm e}^{\Delta_i y}$. 
A choice for the boundary condition is that there are no
poles in the right half plane and thus no supercritical 
contribution originating from the region of small $u$. 
The singularities of $\bar{f}(\omega,u_0)$ are now to the left
of $\omega=0$ and thus not important, and the hard part of 
the Pomeron dominates.
In that case the dominant part in the limit $\beta\sim 0$ is 
contained in the function $\bar{K}_r(\omega,u)$ of Eq.(\ref{Melliny}).
Such an approach which is only based on the $\bar{K}_r(\omega,u)$ 
and not on boundary conditions could in the past provide a 
reasonable description of the first HERA data. 
The present low $x$ data show the need of a supercritical 
intercept input.
The Mellin transformation from $\omega$ to $y$ can be performed and
yields a modified Bessel function $I_1$.
Thus one finds for $\bar{K}_r(y,u)$ after inverse transformations 
over $\omega$ and $\beta$
\bel{Greenrun}
{K}_r(y,u)=\frac{1}{2\pi i}\int_{d-i\infty}^{d+i\infty}
d\beta~{\rm e}^{u\beta}\,
\sqrt{\left(\frac{\tilde{\alpha}_s{\cal R}(\beta)}{y}\right)}
I_1(2\sqrt{y\tilde{\alpha}_s{\cal R}(\beta)})~~.
\ee
For large rapidities one may use $I_1(y)\simeq {\rm e}^y/
\sqrt{2\pi y}$ so that Eq.(\ref{Greenrun}) simplifies to
\bel{Greenrunly}
{K}_r(y,u)=\frac{1}{2\pi i}\int_{d-i\infty}^{d+i\infty}
d\beta~ 
\left(\frac{\tilde{\alpha}_s{\cal R}(\beta)}{16\pi^2 y^3}
\right)^{1/4}
\exp{\left(u\beta+2\sqrt{y\tilde{\alpha}_s{\cal R}
(\beta)}\right)}~~.
\ee
By using the method of steepest descent, the inverse Mellin 
transformation then yields the standard double logarithmic 
approximation to the DGLAP equation  
\bel{APdoublog}
{K}_r(y,u)\sim u^{-1} \left(\frac{\tilde{\alpha}_s
\log u}{16\pi^2 y^3}\right)^{1/4}
\exp\left(\sqrt{4\tilde{\alpha}_s y\log u} \right)~~.
\ee
This solution corresponds to the well known double logarithmic
approximation where not only a strong monotonous increase of 
rapidity is imposed, but also a strong ordering of the transverse 
momenta along the ladder.
It thus means that the resulting cross section grows as 
${\rm e}^{c \sqrt{y}}$, slower than for the observed 
``supercritical Pomeron'', which behaves as 
${\rm e}^{y\Delta_{exp}}$. 

The above boundary condition seems not to be realistic if one 
tries to decribe the soft high energy processes, by assuming 
the existence of only one Pomeron, containing both soft 
and hard mechanisms.
In the case where we expect the existence of a supercritical 
Pomeron ($\alpha_P(0) = 1 +\Delta_0 > 1$) coming from QCD 
interactions in the nonperturbative region, we can choose as 
boundary condition
\bel{Pole1}
\bar{f}(\omega,u_0)=\frac{\bar{f}_0}{\omega-\Delta_0}~~.
\ee
Since $\Delta_0>0$ is to the right of the essential 
singularity of $\bar{K}_r(\omega,u)$, the boundary condition 
gives the dominant contribution at large $y$ while the
singularity of $\bar{K}_r(\omega,u)$ at $\omega=0$ represents the 
hard corrections.
(Strictly speaking, at the boundary $u = u_0$ the behaviour of 
the gluon density may have to be chosen also in accordance with 
the BFKL equation because, depending on $u_0$, the perturbative 
dynamics contained the BFKL equation may already apply.)
To analyse the influence of the soft, low transverse momentum 
region in the BFKL equation one can introduce a lower cut-off 
in the integrals over $k_{\perp}$ in the BFKL equation. 
But it is more complicated to apply this method to the case of 
a running $\alpha_s$ than for a fixed one \cite{ColLan1,MFR}.
In the next sections, we will discuss in more detail the physical 
picture of the gluon ladder for this choice of boundary condition
corresponding to supercritical behaviour.
If one supplies the boundary condition (\ref{Pole1})
the mean transverse momenta in inelastic processes will not
grow with energy; this will also be seen later. 

The main result of this section is that the hard part of the BFKL
Pomeron for the case of a running $\alpha_s$ leads to a cross
section behaving as ${\rm e}^{c \sqrt{y}}$, growing slower with $y$
than the usual ``supercritical'' Pomeron. 
The situation where the leading $y$ dependence is found 
to be determined through the boundary condition, rather 
than the BFKL kernel, will be discussed below.

\section*{\bf 4. The diffusion approximation to the BFKL kernel
 with running $\alpha_s$}
\renewcommand{\theequation}
           {4.{\arabic{equation}}}\setcounter{equation}{0}
We now discuss two main aspects which determine the character of 
a parton cascade represented by our effective gluon ladders.
They are the branching of the gluons leading to an exponential
growth of their number with rapidity and the gluon diffusion
in the transverse momentum plane.
Both determine how the cascade evolves from the fragmentation region
of a fast particle towards ``wee'' partons. 
Of particular importance is whether the branching and diffusion 
velocities depend on $k_{\perp}$. 
In asymptotically free theories the branching velocity becomes 
small for high $k_{\perp}$.
In this case all properties of the parton distribution at large
$k_{\perp}$ will be determined by the relative weight of the
diffusion and the DGLAP-evolution of partons towards larger $k_{\perp}$.
The scale invariant BFKL equation with {\it fixed} $\alpha_s$ has 
the unique property that the splitting probabilities of partons 
do not depend on $k_{\perp}$.
Therefore the distribution of partons in $k_{\perp}$ will be determined
entirely by the initial distribution at the end of the ladder, 
which for large $y$ is spread out over an increasingly larger 
interval of $k_{\perp}$ by the diffusion mechanism.
In this section we will study these two aspects for the BFKL Pomeron 
for running $\alpha_s$.

\subsection{The diffusion-like part in the operator equation}
Due to its convolution structure, it is possible to rewrite the BFKL 
equation Eq.(\ref{BFKLeqrun2}) as
\bel{bfkl_op}
\frac{1}{\alpha_s(u)}\frac{\partial h(y,u)}{\partial y}=\frac{3}{\pi}
{\cal L}(\hat{\beta})h(y,u)~~.
\ee
Here $\hat{\beta}$ is an operator acting in the space of functions
depending on $u$.
It follows from the form of the Mellin transformation
that the operator is conjugated to $u$: $\hat{\beta} = 
\frac{\partial}{\partial u}$.
A useful property of the operator kernel ${\cal L}$ is that 
it satisfies the ``shifting relation''
\bel{kernel_opshift}
{\rm e}^{c u}{\cal L}(\hat{\beta}+c){\rm e}^{-c u}=
{\cal L}(\hat{\beta})~~,
\ee
which can be proven by showing that both sides have identical
Mellin transforms, $\tilde{\cal L}({\beta})$.
As a consequence the ``shifted'' functions $h_{c}(y,u)={\rm e}^{cu/2}
h(y,u)$ obey the differential equation
\bel{bfkl_opshift}
\frac{1}{\alpha_s(u)}\frac{\partial h_c(y,u)}{\partial y}=
\frac{3}{\pi}{\cal L}(\hat{\beta}+c/2)h_c(y,u)~~.
\ee
Because we know the dependence of the BFKL integral
operator on the variable $\beta$ through $\tilde{\cal L}$ given by Eq.(\ref{kernel}), it is possible to find a diffusion approximation.
The function $\tilde{\cal L}(\beta)$ has a minimum at
$\beta=1/2$ as can be seen from Eq.(\ref{kern.exp}). 
For Eq.(\ref{bfkl_opshift}) it means that the first order derivative
with respect to $u$ vanishes when the argument of ${\cal L}$
equals $1/2$.
We therefore choose  $c = 1$ in Eq.(\ref{bfkl_opshift}) and
expand in $\hat{\beta}$.
We introduce the function $ h_1(y,u)={\rm e}^{-u/2}h(y,u)$
and neglect third and higher order derivatives,
which leads to the differential equation
\bel{bfkl_dif}
\frac{1}{b\alpha_s(u)}\frac{\partial  h_1}{\partial y}
\simeq\tilde{\Delta} \,  h_1 + \tilde{B}
\frac{\partial^2  h_1}{\partial u^2}~~~~;~~
\tilde{\Delta}=4\tilde{\alpha}_s\log(2)~~,
~\tilde{B}=14\tilde{\alpha}_s\zeta(3)~~.
\ee
This equation resembles a diffusion equation with branching, but with
a diffusion coefficient $\alpha_s(u) b\tilde{B}$ and a branching 
coefficient $\alpha_s(u) b\tilde{\Delta}$. 
Here the rapidity $y$ plays the role of time and $u$ that of place.
This analogy will be useful for understanding the structure 
of the function $h(y,u)$.  
The diffusion coefficient can be interpreted as a diffusion
velocity, determining how fast the gluons diffuse in $y$.
Since it is proportional to $\alpha_s(u)$, the gluons 
diffuse slowly to large $u$; the region $\beta \sim 1/2$
corresponds to an evolution where changes of $u$ in 
individual steps are not large.
Eq.(\ref{bfkl_dif}) must be supplemented with some 
boundary conditions at small $u=u_0$, representing infrared
low $k_{\perp}$ physics not contained in Eq.(\ref{bfkl_dif}).
 
For the case with fixed $\alpha_s$, Eq.(\ref{bfkl_dif}) 
is a standard diffusion equation with branching,
\bel{difus}
\frac{\partial  h_1}{\partial y}  =
\Delta\, h_1 + B\,\frac{\partial^2  h_1}{\partial u^2}~~,
\ee
where the branching and diffusion coefficient are independent of $u$.
The solution $h(y,u)={\rm e}^{u/2} h_1(y,u)$ is then given by 
Eq.(\ref{sol1}) with the Green's function (\ref{Greenfix}). 
It should therefore be clear from Eq.(\ref{difus}) that for the 
diffusion behaviour predicted by BFKL scale invariance is essential.

We have seen above that the region $\beta\sim 1/2$ corresponds to 
an evolution where changes of $u$ in individual steps are not large: 
${\delta} u \sim \alpha_s$.
On the other hand, large jumps in $u$ are related to the region 
$\beta\sim 1/u\rightarrow 0$, where $\tilde{\cal L}(\beta) 
\sim 1/\beta$ and for which ${\delta} u \sim \alpha_s u$.
In this case we have to set $c=0$ and expand around $\beta=0$; this 
will be done in the next section.

With our choice for $\alpha_s(u)$, Eq.(\ref{runningalpha}), 
the differential equation (\ref{bfkl_dif}) becomes
\bel{de}
 u\,\frac{\partial  h_1}{\partial y}
= \tilde{\Delta}\,  h_1 + 
\tilde{B}\,\frac{\partial^2  h_1}{\partial  u^2}~~.
\ee
A solution of Eq.(\ref{de}) can be simply found by performing
a Laplace transformation with respect to $y$.
For the transformed function ${\bar h}(\omega,u)$ it follows that it 
satisfies the differential equation
\bel{de_o}
u\,\omega \,{\bar h}(\omega,u)=
\tilde{\Delta}\,{\bar h}(\omega,u) 
+\tilde{B}\,\frac{\partial^2 {\bar h}(\omega,u)}{\partial u^2}~~,
\ee
where for simplicity and without loss of generality for the present
discussion we took $h(0,u) = 0$.
By using the substitution $u=z (\tilde{B}/\omega)^{1/3}+
\tilde{\Delta}/\omega $ the differential equation 
(\ref{de_o}) transforms into the Airy equation, 
$f''(z)- z f(z) =0$, which has the solution
\bel{airy1}
 {\bar h}(\omega,u)={\rm Ai}(z)~~;~~~~~~z=\left(
\frac{\omega}{\tilde{B}} \right)^{1/3}\left(  u- 
\frac{\tilde{\Delta}}{\omega}\right)~~,
\ee
where ${\rm Ai}$ is the Airy function that is finite at 
$z\rightarrow \infty$.  
We note that ${\rm Ai}$ is related to the modified Bessel
function ${\rm K}_{1/3}$ through ${\rm Ai}(z)=(1/\pi)
\sqrt{(z/3)}\,{\rm K}_{1/3}(2z^{2/3}/3)$.

Before we give a full analysis of the solution (\ref{airy1}) in the 
next section, we first examine the dominant behaviour 
at large $u$.
If we assume a boundary condition for $ h_1(y,u)$ at small
$u$ of the form
\bel{fg}
  h_1(y, u_0 \sim 1)\sim f_0 {\rm e}^{\Delta_0 y}~~,
\ee
then for $u \gg 1$ one can try to find a solution of Eq.(\ref{de}) 
in the form
\bel{solrun}
 h_1(y,u) \sim \exp{(\Delta_0 y - C_0   u^s - 
C_1   u^{s-1} - \ldots)}~~.
\ee
Upon inserting Eq.(\ref{solrun}) into Eq.(\ref{de}) and comparing the 
leading terms in $  u$ which are proportional 
to $u h_1(y,u)$, we find that $s = 3/2$ and 
$\Delta=\tilde{B} C_0^2\,s^2$. 
Furthermore, comparison of terms proportional to $ h_1(y,u)$ gives 
$\tilde{\Delta}= 2 s(1-s)\tilde{B} C_0 C_1$. 
Solving for $C_0$ and $C_1$ results in
\bel{par_sol2}
s= 3/2~~ ,~~~~~~ C_0 = \pm \frac{2}{3}
\left( \frac{\Delta_0}{\tilde{B}}\right)^{1/2} ~,~~~~~~
C_1 = \mp\frac{\tilde{\Delta}}{\tilde{B}}
\left(\frac{{\Delta}_0}{\tilde{B}} \right)^{-1/2}~~.
\ee
We have two solutions, positive and negative. 
Because a fast increase in $u$ must be rejected, only the
positive solution is accepted for $C_0$
\footnote{It is interesting that if QCD was not asymptotically
free (for example due to a large number of fermions), then the
diffusion in $u$ described by Eq.(\ref{de}) would go in the 
opposite direction of large $u$.
In such a situation only the theory with a fixed limit makes
sense when $\alpha_s(u) \rightarrow \alpha_{max}$ for 
$u \rightarrow\infty$.
So in this case the diffusion in $u$ will move gluons
in the gluonic ladder to higher $u$ and thus the Pomeron
will develop according to the BFKL solution (\ref{sol1}) 
with constant $\alpha_s=\alpha_{max}$.}.
An important point is that the hard Pomeron disappears in
the case of running $\alpha_s$ in this approximation, because the
residue belonging to the pole at $\omega = \Delta_0$ rapidly 
decreases proportional to $\exp{(-C_0 u^{3/2})}$ on a scale 
$u\sim C_0^{-2/3} \sim 1$. 
(The scale for the disappearance of this solution is set by 
$\Delta_0$; the closer it is to the critical value $\Delta_0 = 0$, 
the longer the hard part of the Pomeron survives.)
A significant large $u$ contribution can therefore only come from 
the region $\beta \rightarrow 0$ in $\tilde{\cal L}(\beta)$, 
which will be discussed later. 

The result (\ref{solrun}) can be explained in a simple fashion
that shows how the transverse momentum in the gluon cascade 
evolves as function of rapidity $y$ in the limit $u \gg 1$.
Assume that starting from the soft 
({\em i.e.} small $u$) end of the gluon ladder the gluons
are in the strong coupling region up to rapidities $y-y_1$;
in this region the gluon distribution should then be proportional to
$h_{soft}(y-y_1,u)\sim {\rm e}^{\Delta_0(y-y_1)}$.
After that, in the remaining part of the rapidity interval 
$y_1$, the gluons reach large virtualities $u$ on 
the other end through the diffusion-like mechanism.
This last stage is described by the solution of Eq.(\ref{de}), 
which for large $u$ becomes
\bel{green_u}
h_{hard}(y,u)\sim \exp{(\tilde{\Delta}y/u -u^3 / 9 {\tilde{B}} y)}~~.
\ee
Using this function, we can obtain the gluon distribution
for the full $y$ range through the integral
\bel{int_u_g}
 h_1(y,u)\sim\int_0^y dy_1\,{\rm e}^{\Delta_0(y-y_1)} 
h_{hard}(y_1,u)~~.
\ee
This integral can be estimated at large $y$ by the saddle point
method which gives
\bel{approx_int}
 h_1(y,u)\sim\exp{ \left[ \Delta_0 y- u^{3/2}\frac{2}{3}
\left(\frac{\Delta_0}{\tilde{B}}\right)^{1/2} \right] }~~,
\ee
coinciding with the main term in Eq.(\ref{solrun}).
In Eq.(\ref{int_u_g}) the mean $y_1$ is of the order
\bel{mean_u3}
<y_1>\,\sim\, u^{3/2}\,\frac{1}{3\sqrt{{\tilde{B}}\Delta_0}}~~.
\ee
This shows that at large $y$ the hard part of the Pomeron is
concentrated on its end.
But these statements about the structure of the end part of 
the gluon ladder given by the expressions Eq.(\ref{solrun}) and Eq.(\ref{mean_u3}) do not concern the dominant term, 
since its residue is very small.

\subsection{The infinite sequence of poles}
In this subsection we further study the solution (\ref{airy1})
of the diffusion equation, not restricting ourselves to 
large $u$. 
We pay special attention to the pole structure in the
variable $\omega$, related to the angular momentum through
$j = \omega -1$, and the influence of the boundary conditions.
Our approach, which seems natural for the BFKL and the DGLAP 
equations, is to consider solutions only at $u \ge u_0 $, 
when $u_0$ is not small but also not very far from the 
infrared region.
On the line $u = u_0$ a boundary condition representing the soft 
physics must be given. 
Of course, it is not possible to choose a $u_0$ where {\em only} 
soft physics enter at the boundary; depending on  $u = u_0$ a 
varying amount of hard physics will already be contained in
the boundary condition.

In the variables $\omega$ and $u$, we thus obtain as solution of 
Eq.(\ref{de_o}) 
\bel{sol_con1}
 \bar{h}_1(\omega,u)= \bar{h}_1(\omega,u_0)
\frac{{\rm Ai}(z)}{{\rm Ai}(z_0)}~~,
\ee
where $ \bar{h}_1(\omega,u_0)$ is the boundary condition.
The arguments of the Airy functions $z = z(\omega,u)$ and 
$z_0 = z(\omega, u_0)$ depend on $\omega$ and $u$ as given in 
Eq.(\ref{airy1}).
From the point of view of Regge phenomenology it is obvious
to assume that $\bar{h}_1({\omega},u_0)$ is represented as a 
sum of Regge poles, some of them having a supercritical intercept,
\bel{poles2}
 \bar{h}_1(\omega,u_0) =
\sum_i \frac{\gamma_i({\omega, u_0})}{\omega -\Delta_i}~~.
\ee
As said above, in addition to singularities 
coming from soft physics, some poles contained in the boundary 
condition, Eq.(\ref{poles2}), are connected to the hard physics
contained in the BFKL diffusion kernel. 

Poles in Eq.(\ref{sol_con1}) can also arise from zeros of the 
denominator.  
These additional poles occur when the argument $z$ of the Airy 
function in Eq.(\ref{airy1}) is negative.
This provides an upper bound on the spectrum 
of the singularities in $\omega$ arising from the Airy function,
\bel{upperbound}
\omega_{max} < \frac{\tilde{\Delta}}{u_0}=
4\bar{\alpha}_s(u_0)\log(2)~~.
\ee
We see that the most right singularity in the complex angular
momentum plane, $j$, is smaller than the BFKL intercept 
calculated with the coupling constant at the boundary scale $u_0$.
The position of the zeros of the Airy function are with 
high precision given by the formula \cite{AbraSte} 
$z=-(3\pi n/2-3\pi/8)^{2/3}$ with $n$ a positive integer. 
We can find an approximate value for $\omega_{max}$ by inserting
$\omega_{max} = \tilde{\Delta} / {u_0} - \epsilon$ into the 
expression for $z(\omega,u)$. 
Solving for $\epsilon$, we find
\bel{ommax}
\omega_{max} = \frac{\tilde{\Delta}}{u_0} \left[ 
1-\frac{1}{u_0^{2/3}}
\left(\frac{9\pi}{8\kappa}\right)^{2/3}\right] ~~,
\ee
with $\kappa=(\tilde{\Delta}/\tilde{B})^{1/2}\approx 0.47$.
It is also possible to obtain the locations of the poles
in the limit $\omega \ll {\tilde{\Delta}}/u_0$ where $z_0 \simeq
-({\kappa\tilde{\Delta}}/{\omega})^{2/3}$.
The zeros in the denominator then result in poles located at
\bel{pole_pos}
\omega \simeq \omega_n = \tilde{\Delta}\kappa\left[\frac{3}{2}\pi n
-\frac{3}{8}{\pi}\right]^{-1}~~.
\ee
This shows that in the diffusion approximation with running coupling
constant, the solution contains an infinite set of
poles accumulating at $\omega = 0$ for $n \rightarrow \infty$.
Such a sequence of poles was already found in the 
$\lambda \phi^3_6$-model in Ref.\cite{Lovelace}.
In QCD this sequence of poles was first found by Lipatov \cite{Lip2},
starting from the same generalization of BFKL to a running gauge 
coupling, but using different techniques.
The discussion of these poles in Ref.$\cite{Levin2}$ is more
close to ours.

To clarify the importance of the poles $\omega_n$, we calculate 
their contribution to $ h_1(y,u)$.
Performing the inverse Laplace transformation over $\omega$ we 
obtain
\bel{ycon1}
 h_1(y,u) =\sum_i 
{\rm e}^{y \Delta_i}\lambda(\Delta_i,u) +\sum_{n=1}^{\infty} 
{\rm e}^{y \omega_n}\lambda(\omega_n,u)~~ ,
\ee
where the first term is the contribution from the (mainly soft)
singularities $\Delta_i$ of Eq.(\ref{poles2}) and the second that
from the Airy function in the denominator; $\lambda(\eta,u)$ 
denotes the residue of a pole at $\omega=\eta$.
The residues of the second term, $\lambda(\omega_n,u)$, initially 
oscillate in $u$ and can be large up to $u \sim 3\pi  n/2\kappa$. 
After that, for $u \gg u_0$ and large $n$, they
are proportional to
\bel{pole_res1}
 \lambda(\omega_n,u)\sim 
{\rm Ai}\left(\left[\frac{2}{3\pi n}\right]^{1/3}
\kappa u- \left[\frac{3\pi n}{2}\right]^{2/3}\right)
\sim \exp{\left( -\frac{1}{\sqrt{\pi n}} 
\left(\frac{2\kappa u}{3}\right)^{3/2}\right)}~~ .
\ee
We see that for large $n$, where $\omega_n \rightarrow 0$, 
the residues increase and they become ``harder''. 
Because the residues go quickly to zero, only the largest
$n$ values are essential at very high $u$. 
The $u$ dependence of the residue of a pole at $\omega_n$ 
scales with ${\kappa}^{-1} n^{1/3}$. 
Thus singularities in the sequence $\omega_n$ more to the right 
({\it i.e.} to larger values) are softer.

 
For large $u$, it is simple to estimate the contribution from large $n$
to $ h_1(y,u)$,
Eq.(\ref{ycon1}):
\bel{qqq}
  h_1(y,u) ~\simeq~ \sum_n^{\infty}~ \widetilde{\phi}_n(u)
 \,\exp{\left( -
 \frac{1}{\sqrt{\pi n}} 
\left(\frac{2\kappa  u}{3}\right)^{3/2}+
\frac{2\tilde{\Delta}{\kappa}}{3\pi n} \right) }~~,
\ee
where $\widetilde{\phi}_n(u)$ are weakly dependent on $u$.
The largest term in this series is at the point $n=(6\tilde{\Delta}^2/\pi \kappa)\, y^2/  u^3$.  
Hence the sum can be estimated as
\be
 h_1(y,u)  \sim  \exp{\left( y\tilde{\Delta}/{  u}-
  u^3 \tilde{\Delta} /9\tilde{B}y\right)}~~,
\ee
which is similar to the solution, Eq.(\ref{green_u}). 
So we see that the sum of poles from the BFKL kernel at high $u$ 
can be represented by an effective $u$ dependent intercept with 
$\Delta_{eff}(u)= \tilde{\Delta}/u$.

We now comment briefly on the connection between the boundary 
condition, $h_1(\omega, u_0)$ and the BFKL kernel in Eq.(\ref{airy1}). 
As can be seen from Eq.(\ref{upperbound}), choosing a larger starting 
value $u_0$ moves the upper limit $\omega_{max}$ of the spectrum 
of poles in $\omega$ to the left.
In order to lead to the same result for 
$ h_1(\omega, u)$, 
independent of $u_0$, this change in the kernel due to the poles near $\omega_{max}$, has to be compensated by a concomitant change in 
the boundary function $ h_1(\omega, u_0)$. 
For example, a pole at $\omega_k$ can be removed by a 
zero of $ h_1(\omega_k,u_0)$. If, on the other hand, $u_0$ is 
decreased, a pole $\Delta_i$ originally in the boundary function 
will move into the BFKL kernel.
This connection shows that for a stable choice of $u_0$, the boundary
has to be in a region where the BFKL equation, and especially the 
diffusion approximation, is already applicable.

While we have found an infinite set of poles accumulating towards 
$\omega = 0$, for a fixed coupling constant there is a {\it cut} 
in the $\omega$-plane (see Fig 3.). 
It was remarked in Ref.\cite{Lip2} that the sequence of poles 
$\omega_n $ for running $\alpha_s$ in some sense corresponds 
to a cut in the $\omega$-plane for fixed $\alpha_s$.  
But this analogue is incomplete as far as we can see for our case
with $t = 0$ (forward scattering). 
For fixed $\alpha_s$ the right part of the cut corresponds to the 
hard Pomeron with mean $<u> \sim \sqrt{y}$, while for running 
$\alpha_s$ the right poles are relatively soft with $<u>\sim 1$, 
and only the left side of the sequence $\omega_n$ towards 
$\omega =0$ becomes hard.
These poles may even be artifacts due to our particular 
choice of running $\alpha_s$, which allowed an analytical solution. 
Furthermore, they may disappear when other higher order corrections 
to the Pomeron, for example the mixing with multiguon states in the 
$t$-channel, are taken into account. 
These questions go beyond the scope of our paper and need more 
investigation.  

We can get a slightly different view on the evolution of 
the gluon ladder and the structure of the $\omega_n$-poles 
by considering a quantum mechanical analogy. 
Eq.(\ref{bfkl_dif}) can be seen as the Schr\"{o}dinger equation for
a one-dimensional motion in the coordinate $u$ in the interval 
$u_0 < u < \infty$. 
The energy of this motion --- when time is identified with
$t=-iy$ --- is given by
\bel{efham}
E=\frac{1}{2 m(u)}{\hat p}^2+V(u)~~,
\ee
where we defined the momentum, potential and coordinate dependent mass, respectively, as
\bela{mpc}
{\hat p}~~&=&-i {\hat \beta} = -i\partial/\partial u~~,
\\V(u)&=&-\tilde{\Delta}/u~~, 
\\m(u)&=& {u}/{2\tilde{B}}~~.
\eela
In the case of a constant $\alpha_s$ we have free motion.  
So when we put the ``particle'', represented by a wave packet, 
at an initial time at some position $u_1$, then the situation 
at a later time is described by the Green's function for free motion, Eq.(\ref{sol2}), yielding propagation and spreading of the wave packet.
But for running $\alpha_s$ the motion is not free; we have a
long ranged attractive force in the direction to smaller
``distances'' $u$. 
In this potential an infinite ``Coulomb-like'' series of bound states 
exist. 
For the high lying states, the particle is located on the average at 
large distances $ u \gg u_0$ where the motion is quasiclassical.
To find their energy spectrum, $E_n$, we use the standard 
quasiclassical quantization condition
\bel{qcon}
n\pi =\int_{u_0}^{u_{max}(n)} d u\,p(u) =
\int_{u_0}^{\tilde{\Delta}/E_n} d u\, \sqrt{\left[\frac{\tilde{\Delta} - 
E_n u}{\tilde{B}}\right]}\simeq \frac{2}{3} 
\sqrt{\left(\frac{\tilde{\Delta}}{\tilde{B}}\right)}  \frac{\tilde{\Delta}}{E_n}~~,
\ee
which leads to $E_n =(1/n)\cdot(2\tilde{\Delta}^{3/2}/3\pi
\tilde{B}^{1/2})$, coinciding with Eq.(\ref{pole_pos}) in the limit
of large $n$.
It is interesting to note that the
behaviour of the residues in Eq.(\ref{pole_res1}) corresponds to the 
motion of our ficticious particle in the classically forbidden 
region $u> \tilde{\Delta}/ E_n$.
The momenta in this region are not constant due to the $u$ dependence
of the mass,  
\bel{momentum}
p_n(u) =\sqrt{-2 m(u) E_n} = i\sqrt{\frac{2\kappa^{3}}{3\pi}
\frac{u}{n}}
~~.
\ee
Using the wave function $\lambda_n(u) \sim \exp{(i u p_n(u))}$, 
we obtain the same result as in Eq.(\ref{pole_res1}).

So far we have excluded the particle from the region $u < u_0$,  
corresponding to an infinite potential barrier. We can change the 
potential (and thus the boundary condition at $u_0$) in order 
to allow the ``particle'' to go to smaller $u$.
This can for example be achieved if we choose the potential
\bel{pot.dif}
  V(u) =-\frac{\tilde{\Delta}}{\log{(v(u) + {\rm e}^u)}}~~,
\ee
where the function $v(u) \rightarrow 0$ at $u \rightarrow \infty$
and $v(u) \rightarrow v_0 >1$ at $u \rightarrow -\infty$.  
At large $u > 0$ we have the same potential as before
but now we can put $u_0 = -\infty$ .  
In this case the motion in $u < 0$ will be 
unlimited also for some $E < 0$, and all discrete eigenvalues 
$E_n$ disappear. 
In addition, we can choose $v(u)$ in such a way that 
$V(u)$ will have a local minimum at $u \sim 1$ so that we 
generate a low lying bound state.  
In terms of our pole structure in $\omega$, this corresponds to a 
``soft'' state  with a high intercept resulting from this
potential.

In Ref.\cite{KirLip} the structure of the BFKL Pomeron for $t < 0$
was studied. 
It is amusing that we can arrive at their findings concerning the 
Pomeron trajectory by using the following shortcut based on our
quantum-mechanical analogue.
The position of the Pomeron pole at some momentum transfer 
$t < 0$ can be simulated from our $t = 0$ example by restricting 
the range of the ``coordinate'' $u$ 
({\it i.e.} the range of the transverse momenta, through the 
condition $u > u_0 = \log{(-t)}$.
We choose in Eq.(\ref{qcon}) a large 
$u_0$ and simply put $n=1$ to obtain the leading BFKL Pomeron 
trajectory,
\bel{p-traj}
\omega_P(t)\,\simeq\, E_1 \,\simeq\, 
\frac{\tilde{\Delta}}{u_0(t)}~~,
\ee
where $u_0= \log{(-t/\Lambda^2)} \gg 1$.
The trajectories for the other ``satellite'' poles also can 
be estimated for large $u_0$ from Eq.(\ref{qcon}),
\bel{sat-traj}
\omega_n (u_0) \simeq {\Delta}(u_0)\left[1-
\left(\frac{3\pi}{2 \kappa}
\frac{n}{u_0} \right)^{2/3} \right]~~.
\ee
These relations for the Pomeron trajectory coincide with 
the ones that were found in Ref.\cite{KirLip} in a precise 
(and therefore more complicated) way.
The above results also show the explicit dependence on the infrared
cut-off $u_0$.
 
In conclusion, we repeat that the general significance of 
the poles at $\omega_n$ needs future investigation. 
They may be a feature closely connected to our particular choice 
of the running coupling contant. 
It will be shown below that for large $u$ their contribution 
is small in comparison with the $\beta \sim 0$ part of the
kernel, corresponding to the DGLAP mechanism.

\section*{5. Pomeron with hard ends}
\renewcommand{\theequation}
           {5.{\arabic{equation}}}\setcounter{equation}{0}

Eq.(\ref{bfkl_op}) has the structure of the kinetic equation. 
The differential equation (\ref{bfkl_dif}) discussed above
corresponds to the diffusion approximation,  
where fast increases in the ``coordinate'' $u$ do not occur. 
For the BFKL Pomeron with fixed $\alpha_s$ such jumps in 
$k_{\perp}$ are not essential, because the splitting of 
gluons is universal for all $k_{\perp}$ and the diffusion mechanism is 
sufficient to evolve the gluon distribution in the ``time" $y$ to
obtain $f(y,u)$ at larger $u$. 
All possible inhomogeneities, arising from initial conditions or the 
infrared region, are smeared out by diffusion.

As we saw above, for the BFKL Pomeron with running $\alpha_s$ the 
situation is different. 
In the diffusion approximation, for which the vicinity 
around $\beta = 1/2$ of $\tilde{\cal L}$ is essential, the gluon 
splitting probability ${\tilde B} \alpha_s (u)$ at large 
$k_{\perp}$ decreases with $u$. 
Therefore only a small fraction of the initially moderate 
$k_{\perp}$ gluons diffuses to the region of larger $k_{\perp}$.
As result the density $f(y,u)\sim u {\rm e}^{-u/2}h(y,u)$  
decreases fast with $u$. 
This has been discussed in detail above. 

In contrast to the diffusion-like behaviour resulting from 
the expansion of the BFKL kernel $\tilde{\cal L}$ around its minimum,
the limit $\beta \rightarrow 0$ of the kernel leads to a growth of  
$f(y,u)$ as function of $u$ as shown in Eq.(\ref{APdoublog}).
For a BFKL Pomeron with running $\alpha_s$ only this part can generate
a significant contribution at large $u$. 
In this section we briefly show how these previous
results, the low $x$ DGLAP solution, Eq.(\ref{APdoublog}), 
follow quite directly from the BFKL operator equation 
Eq.(\ref{bfkl_op}).

It is well known that a DGLAP-like evolution equation follows 
simply from BFKL when one takes into account only configurations 
strongly ordered in $k_{\perp}$; this limiting case corresponds to 
the part of the eigenvalue spectrum $\tilde{\cal L}(\beta)$ situated 
at $\beta  \rightarrow 0$, where $\tilde{\cal L}(\beta)\sim 
1/\beta $.
In the operator equation (\ref{bfkl_op}), the rapidly increasing 
part of the BFKL Pomeron is found by using the approximation 
to the BFKL operator  
\bel{oper2}
{\cal L}(\hat{\beta}) \simeq \hat{\beta}^{-1}=
\int^u d u' \left[\ldots\right]~~. 
\ee
Thus instead of Eq.(\ref{bfkl_dif}) we get the integral equation
\bel{bfkl_ap}
\frac{1}{b\alpha_s(u)}\frac{\partial  h_1(y,u)}{\partial y} =
\tilde{\alpha}_s\left[\int_{u_0}^{u} d u'\,  h_1(y,u')~+~ 
 h_1(y,u_0)\right]
~.
\ee
This approximation is plausible if one looks at
Eq.(\ref{BFKLeq}):  take $k_{\perp}\gg q_{\perp}$ in the integral of Eq.(\ref{bfkl_op}), which then transforms  directly into 
Eq.(\ref{bfkl_ap}) if we introduce the same integration limits.
Eq.(\ref{bfkl_ap}) can be solved by means of a Laplace transformation
in $y$ and taking the derivative with respect to $u$, which yields
\bel{sol_bfkl_ap}
 h_1(y,u)=\int_{c-i\infty}^{c+i\infty}d\omega\, {\rm e}^{\omega y}
 \bar{h}_1(\omega,u_0)\,\,\left(\frac{\alpha_s(u_0)}
{\alpha_s(u)}\right)^{-1+\tilde\alpha_s/\omega}
~~,
\ee 
which is the same DGLAP asymptotic solution as in Eq.(\ref{APsol1}).

From our considerations follows a simple model for the Pomeron
suitable for cases when one probes the Pomeron in a hard
process, as in deep-inelastic scattering.
In fact it is prompted by Eq.(\ref{sol_bfkl_ap}), where 
${h}_1(y,u)$ is written as the inverse Laplace transformation
of the product of two functions: 
the first function is the boundary condition function 
$\bar{h}_1({\omega},u_0)$, containing mainly the soft, 
non-perturbative physics, and the second one is the BFKL kernel 
with running $\alpha_s$, $\bar{K}_{hard}(\omega,u)$, 
containing the essential singularity in $\omega$.  
For the boundary condition we expect a supercritical 
behaviour and therefore use the function given in 
Eq.(\ref{Pole1}).
We use this structure to directly model the unintegrated 
gluon distribution $f(y,u)$.
For our model we choose 
\footnote{A similar expression was already proposed in 
Ref.\cite{AndGus1} in the context of the linked dipole chain 
model \cite{AndGus2}.
In Ref.\cite{AndGus1} it represents an interpolation between 
DGLAP and the hard BFKL Pomeron. 
We thank G.Gustafson for his remarks.}
\bel{soft_ap}
f(y,u) = f_0 \int_0^y d y_1\, {\rm e}^{(y-y_1)\Delta_0}\, 
{K}_{hard}(y_1,u)~~,
\ee
where
\bel{Vv}
{K}_{hard}(y,u) = \frac{1}{2 \pi i} \int_{c-i\infty}^{c+i\infty} 
d\omega\,{\rm e}^{\omega y}\bar{K}_{hard}(\omega,u)~~.
\ee
Since we are interested in a hard process where $u$ is large,
we can model ${K}_{hard}(y,u)$ by a DGLAP-type kernel that contains
all essential features found in Section 3:
\bel{ap_vertex}
{K}_{hard}(y,u)\simeq I_0(2\sqrt{y \xi(u)})~~,
\ee
with $I_0$ the zeroth order modified Bessel function depending
on the ``invariant charge''
\bel{inv.charge}
\xi(u)\equiv \tilde{\alpha}_s
\log{\frac{\alpha_s(u_0)}{\alpha_s(u)}}~~.
\ee
In Eq.(\ref{soft_ap}), the transverse momenta of the gluons are
large up to $y_1$.
The values of $y_1$ where the dominant contribution to the
integral in Eq.(\ref{soft_ap}) comes from, can in general depend 
on the total rapidity $y$.
However, since the $y$ dependence of the integrand factorizes in 
our model, it is obvious that the dominant $y_1$ region 
at very large $y$ only depends on $u$, being independent 
of $y$.
We calculate the dominant interval $y_1$ below.

Expanding the modified Bessel function $I_0$, one obtains
\bel{sap7}
f(y,u) =  f_0\, {\rm e}^{y \Delta_0}\,\frac{1}{\Delta_0}
\sum_{n=0}^{\infty}
\frac{1}{n!}
\left[ \frac{\gamma(n+1, y\Delta_0)}{n!} \right]
\left(\frac{\xi(u)}{\Delta_0}\right)^n~~,
\ee
where $\gamma(n,x)$ is the incomplete $\Gamma$-function. 

For large $y$, the Bessel function can be approximated as 
\bel{hardkernel}
I_0(2\sqrt{y \xi(u)}) \simeq
\frac{1}{2\sqrt{\pi \sqrt{y \xi}}}
\exp{(2\sqrt{y \xi(u)})}~~,
\ee
and the integral in our model, Eq.(\ref{soft_ap}),
can be extended to $\infty$. 
It then can be done analytically \cite{AbraSte},
\bea
f(y,u) &=& f_0\int_0^y  d y_1\,{\rm e}^{(y-y_1)\Delta_0}
 I_0(2\sqrt{y_1 \xi(u)})  \simeq\,
f_0{\rm e}^{\Delta_0 y} \int_0^{\infty} d y_1\,{\rm e}^{-y_1\Delta_0}
I_0(2\sqrt{y_1 \xi(u)}) \label{sappp}\\
&=& \frac{f_0}{\Delta_0}
{\rm e}^{{\Delta_0 y}+\xi(u)/\Delta_0}= \frac{f_0}{\Delta_0}
{\rm e}^{\Delta_0 y}\,\left(\frac{u}{u_0} \right)^{\frac{3}
{\pi b\Delta_0}}~~ \label{sap2}.
\eea
This answer can also be obtained from Eq.(\ref{sap7}) by 
noting that for $y\rightarrow\infty$ the factors between brackets 
approach $1$ so that Eq.(\ref{sap7}) reduces to Eq.(\ref{sap2}). 
The solution shows, as expected, a $y$ dependence characterized
by the Pomeron intercept $\Delta_0$ that originates from the soft
boundary, and has multiplicative hard corrections, which again 
involve $\Delta_0$. 
For large $y$, the dominant contribution to the above integral can 
be easily shown to come from the region around 
a saddle point at
\bel{meany3}
 y_1^{(s)} = \xi(u) /\Delta_0^2~~.
\ee
The quantity $\Delta_0\sim 0.1$\,--\,$0.3$ is small,
and depends on the chosen boundary condition ({\em i.e.} on the
intercept of the Pomeron at scale $u_0$).
The invariant charge $\xi(u)$ starts to grow relatively fast at 
$u_0$ as can be seen from Eq.({\ref{inv.charge}}).
The subsequent growth becomes very slow due to its doubly logarithmic 
form.
For a boundary scale $k_0^2=2$ GeV$^2$ and $\Lambda=0.2$ GeV,
one obtains $\xi(u)=0.3$ at $k_{\perp}^2=5$ GeV$^2$,  $\xi(u)=1$ at 
$k_{\perp}^2=10^2$ GeV$^2$ up to $\xi(u)=1.7$ at a very large scale, 
$k_{\perp}^2=10^4$ GeV$^2$.
Since the denominator in Eq.(\ref{meany3}) is small,
the saddle point $y_1^{(s)}$ is already large -- of order 10 --
at moderate transverse momenta.

A similar estimate for the relevant rapidities $y_1$ can be 
obtained by calculating the mean $y_1$, weighted with the 
normalized integrand in Eq.(\ref{soft_ap}):
\bel{meany4}
<y_1> = (\xi(u)+\Delta_0) /\Delta_0^2\simeq  y_1^{(s)}~~.
\ee
For the fluctuation around the mean of $y_1$ one finds
\bel{flucy}
  {\delta} y_1=\sqrt{<y_1^2> -  <y_1>^2 }=\sqrt{ \frac{2 \xi(u) +
\Delta_0}{\Delta_0^3}}\simeq\sqrt{\frac{2 y_1^{(s)}}{\Delta_0}}~~,
\ee
which shows that the fluctuations in $y_1$ around its mean are 
proportional to $1/\sqrt{y_1^{(s)}}$ and thus small for 
$y_1^{(s)}\gg 2/\Delta_0$.

In current experiments the available maximum rapidity $y$ is of 
order $10$.
For such energies the saddle point $y_1^{(s)}$ can be large
with respect to $y$ and even exceed $y$.
As a consequence, it is then not correct to extend the integration
range in Eq.(\ref{sappp}) to infinity. 
If $y/y_1^{(s)}$ is small, a more adequate representation of
Eq.(\ref{sap2}) is given by 
\bel{bes_ser}
 f(y,u) = f_0 \sum_{n=0}^{\infty} \left( \frac{y~}{y_1^{(s)}} 
\right)^{n/2}
            I_n(2\sqrt{y \xi(u)})~~.
\ee

Coming back to Eq.(\ref{soft_ap}), we see that our model 
provides us with a prescription how to incorporate the hard 
part of the Pomeron, ${K}_{hard}(y,u)$, in reaction amplitudes 
based on the exchange of Reggeons when $y$ is large. 
In this framework it is simple to calculate other
quantities.
In the amplitude for ``onium-onium'' scattering  with virtualities 
$u_1$ and $u_2$,  the contribution of the Pomeron can be written as
\bel{fdi}
A(y,u_1,u_2) \sim \int_0^{\infty} dy_1 dy_2 dy_3\,{K}_{hard}(y_1,u_1)\, 
f_0{\rm e}^{\Delta_0
y_2} {K}_{hard}(y_3,u_2) \delta(y - y_1 - y_2 - y_3 )~~.
\ee
It corresponds to a Pomeron that contains soft physics, in between 
two hard ends and can be diagrammatically represented by Fig.4. 
In the language of Regge theory, this contribution is an enhanced 
diagram. 
In the limit of $y \rightarrow \infty$ the ends can be considered as 
vertices, 
\bel{ooscat}
A(y,u_1,u_2) = g(u_1)\,f_0{\rm e}^{\Delta_0 y}\,g(u_2)~~,
\ee
where the hard vertices have the asymptotic form
\bel{hardver}
g(u) = g_0 \exp{(\xi(u)/\Delta_0)}~~.
\ee
One can apply the hard probe to the Pomeron not only at
the ends. 
When we calculate the inclusive cross section of the production 
of a hard gluon (jet) with rapidity in the central region, 
we get a result analogous to Eq.(\ref{fdi}).
If we assume for simplicity that $y \rightarrow \infty$, so that
the hard parts become hard vertices as in Eq.(\ref{ooscat}), we obtain
\bel{isp}
A_{incl}(y,y_1,u_1,u,u_2) = f_0^2 g(u_1)\,g(u_2)\,{\rm e}^{y\Delta_0 } \,
\int_0^{y_1} dy_a \int_0^{y-y_1} dy_b\,{K}_{hard}(y_a,u) 
{\rm e}^{-\Delta_0(y_a+y_b)} {K}_{hard}(y_b,u)\,~~,
\ee
where $y_1$ and $u$ are the rapidity and 
$\log(k_{\perp}^2)$ of the measured gluon(jet), respectively. 
 
Finally, we compare the fixed and running $\alpha_s$ cases
for onium-onium scattering, which is usually considered as a 
``laboratory'' for the perturbative BFKL equation. 
For simplicity, we choose similar onia with small transverse sizes
${R_{\perp}^2}$.
The transverse momenta of the gluons at the ends of the ladders are 
then large: $u_1= u_2 \sim \log{\frac{1}{(R_{\perp})^2}}$. 
The coupling constant at these scales is of order $1/u_1$.
Because in our example the two onia are equivalent, the internal 
structure of the Pomeron is symmetric around the middle, where the 
rapidity is half of the large full interval $y$.
Therefore we only consider the first half of the ladder.   
At the beginning of the ladder, for rapidities $y_1 \ll u_1$,
the simple two gluon exchange model for the Pomeron is in fact 
adequate, since there is no need for resumming the $\alpha_s y_1$ 
contributions, which are small in this region.
This holds for both fixed and running coupling constant.
After this stage, for a fixed coupling
constant one enters into the true BFKL regime.
The mean $u$ of a gluon in the Pomeron ladder at rapidity interval 
$y_1$ is first approximately constant and of order $u_1$; due to the 
diffusion the transverse momenta spread out with rapidity, 
$<u-<u>>^2\sim {y_1}$, but this spreading happens 
equally towards lower and larger $u$.
Then at rapidities $y_1\sim u_1^2$ the transverse momenta 
of a substantial amount of gluons can reach small values, $u\sim 1$.
A boundary condition supplied at low tranverse momenta 
then ``prohibits'' the transverse momenta $u$ to diffuse into the 
soft region. As a consequence, the mean transverse momenta in the ladder
start to grow in this region. 
For running $\alpha_s$ the sitation is different; now
the amplitude is given by Eq.(\ref{ooscat}).
The transverse momenta, or $u$, decrease in large steps, but remain 
large up to rapidities $y_1\sim \xi(u)/\Delta_0^2$.
At this point they reach the soft regime.
In the limit of very large $y$ the ends are relatively small
and the soft part covers the most part of the rapidity interval $y$.
 
We thus obtained a model for the Pomeron in hard scatterings
that lends itself to a simple diagrammatic interpretation.
In this model the Pomeron with a running $\alpha_s$
is of soft, nonperturbative nature, but it has hard ends when probed 
in hard scatterings.
The hard ends become small in the limit of very large $y$.
However, in the current deep-inelastic scattering experiments 
the total rapidity intervals are to small to see the soft nature
of the Pomeron since the hard ends are large, thus leading to
large hard corrections.


\section*{6. Conclusion}
\renewcommand{\theequation}
           {.{\arabic{equation}}}\setcounter{equation}{0}
The original derivation of the BFKL Pomeron was carried out 
for a fixed coupling constant in the leading logarithmic 
approximation, $\alpha_s y \sim 1$. 
The physical picture behind it is a gluonic ladder where the 
rungs are strongly ordered in rapidity, the so called 
multiregge kinematics. 
Given a characteristic scale at one end of the ladder, the 
transverse momenta of the gluons diffuse.
The variance of $u = \log(k^2_{\perp})$ is of the order of $y$, the
total rapidity difference along the ladder.
The ``hard'' BFKL Pomeron applies if this diffusion doesn't reach into
the soft nonperturbative regime. 
The gluon splitting along the ladder leads to the exponential 
growth of the gluon density with rapidity, characterized by the BFKL 
intercept $\Delta \sim 2.6\, \alpha_s$. 
This intercept follows from a cut in the complex angular momentum 
plane.

Assuming a standard form for $\alpha_s(k_{\perp})$, we extended the
BFKL equation to the case of a running coupling constant.
We were able to obtain an analytical solution for the gluon 
distribution, transformed to $\beta$ and $\omega$, the conjugate 
variables of $u$ and $y$.  
It has an essential singularity at $\omega = 0$, corresponding to a 
cross section of the form ${\rm e}^{ c \sqrt{y}}$. 
As a consequence, the singularities introduced through the boundary 
conditions at some low scale $u_0$ are crucial, because they give 
the dominant contributions if they are to the right of the 
essential singularity. 
It was checked that our solution is consistent with the 
DGLAP expression at large $y$, corresponding to low $x$.

We expanded our analytical solution for a running coupling constant
around $\beta = 1/2$. 
This leads to a diffusion-like equation, where the diffusion and 
splitting coefficients now depend on $u$.  
This is in contrast to the hard BFKL Pomeron with fixed $\alpha_s$, 
where these coefficients are constant.
In this regime the changes in $u$ along the ladder are relatively
small.
We found in this approximation an infinite sequence of poles
in $\omega$, accumulating at $\omega = 0$. 
The poles most to the left are the hardest, {\it i.e.} most important 
at large $u$.
They can be effectively represented by a $u$ dependent 
Pomeron intercept, going to $0$ for large $u$. 
This behaviour could be explained by considering the semi-classical 
limit of a quantum mechanical analogue. 
We also made a simple estimate for the behaviour of the Pomeron 
trajectory for $t \ne 0$.

In general, the diffusion-like part of the solution was found to
yield a relatively small contribution if the changes in
the transverse momenta are large. 
The important part is then due to the $\beta \sim 0$ region of 
the BFKL operator, where large changes in $u$ occur. 
This is similar to the behaviour  predicted by
the DGLAP equation. 
We proposed a simple model for large $u$, which reflects all the 
essential properties of the exact solution: a Pomeron which is 
of nonperturbative origin, but has hard ends when probed by a 
hard device.
In the limit of large $u$ and large $y$ a diagrammatic 
representation was given, which enables one to calculate Pomeron 
exchange amplitudes quite simply.
Onium-onium scattering was discussed as an example in this 
framework.

In our study we included the running of the coupling constant,
an effect we consider the most obvious source of the breaking 
of scale invariance of the BFKL equation. 
This led to properties of the resulting Pomeron, which are very
different from the characteristics of the perturbative
hard Pomeron which one obtains with fixed $\alpha_s$.
There are of course other higher order corrections which must be
taken into account. 
Examples are corrections to the Lipatov vertices and more 
complicated multi-gluon exchanges. 
Such corrections can still be collected in a modified single Pomeron  
exchange.
There are however also contributions to the reaction amplitude
beyond the single Pomeron exchange. 
They include, for example, the sequential exchange of more 
than one Pomeron and the ``triple Pomeron'' interaction.
If these multi-Pomeron states become very important, for example
at superhigh energies where they are needed to unitarize reaction 
amplitudes, the significance of the single Pomeron as a building 
block will be reduced. 
However, our study has shown that as long as one uses it as a 
building block, the running of the gauge coupling must be included 
as it changes the character of the Pomeron significantly.
\vspace*{1cm}\\
\centerline{\bf  ACKNOWLEDGEMENT}
\vspace*{0.1cm}\\
We want to thank K.Boreskov and A.Kaidalov for numerous 
discussions and comments. 
We are also grateful to V.S.Fadin, V.N.Gribov, G.Gustafson, V.Khoze,
Y.A.Simonov, K.A.Ter-Martirosyan for their interest in this work. 
O.K. would like to thank the theory group of NIKHEF for its
hospitality.
The work of L.H. and J.K. is part of the research program
of the Foundation for Fundamental Research of Matter (FOM)
and the National Organization for Scientific Research (NWO).
The collaboration between NIKHEF and ITEP was supported by a grant 
from NWO and by INTAS grant 93-79.
O.K. also acknowledges support from grant J74100 of
the International Science Foundation and the Russian Government



\section*{\bf Figure Captions}

\begin{figures}{}

\item[{\bf Fig.1}] Contribution of a gluonic ladder with
$n$ rungs to Pomeron exchange.
Lipatov vertices are denoted by dots and reggeized
gluon propagators by thick vertical gluon lines.
\item[{\bf Fig.2}] Solid line: $\tilde{\cal L}(\beta)$.
Dashed line: ${\cal R}(\beta)$.
\item[{\bf Fig.3}] {\bf a.)} For fixed $\alpha_s$: branch cut in 
the $\omega$-plane starting from ${\rm Re}~\omega=\Delta$. 
{\bf b.)} For running $\alpha_s$: series of poles
accumulating at $\omega\rightarrow 0$ starting at 
${\rm Re}~\omega=\omega_{max}$. 
\item[{\bf Fig.4}] Diagrammatic representation of Pomeron 
exchange in onium-onium scattering. Hatched blocks: hard ends.
Middle: soft part of Pomeron.
\end{figures}

\newpage
\centerline{\epsfysize=8cm \epsffile{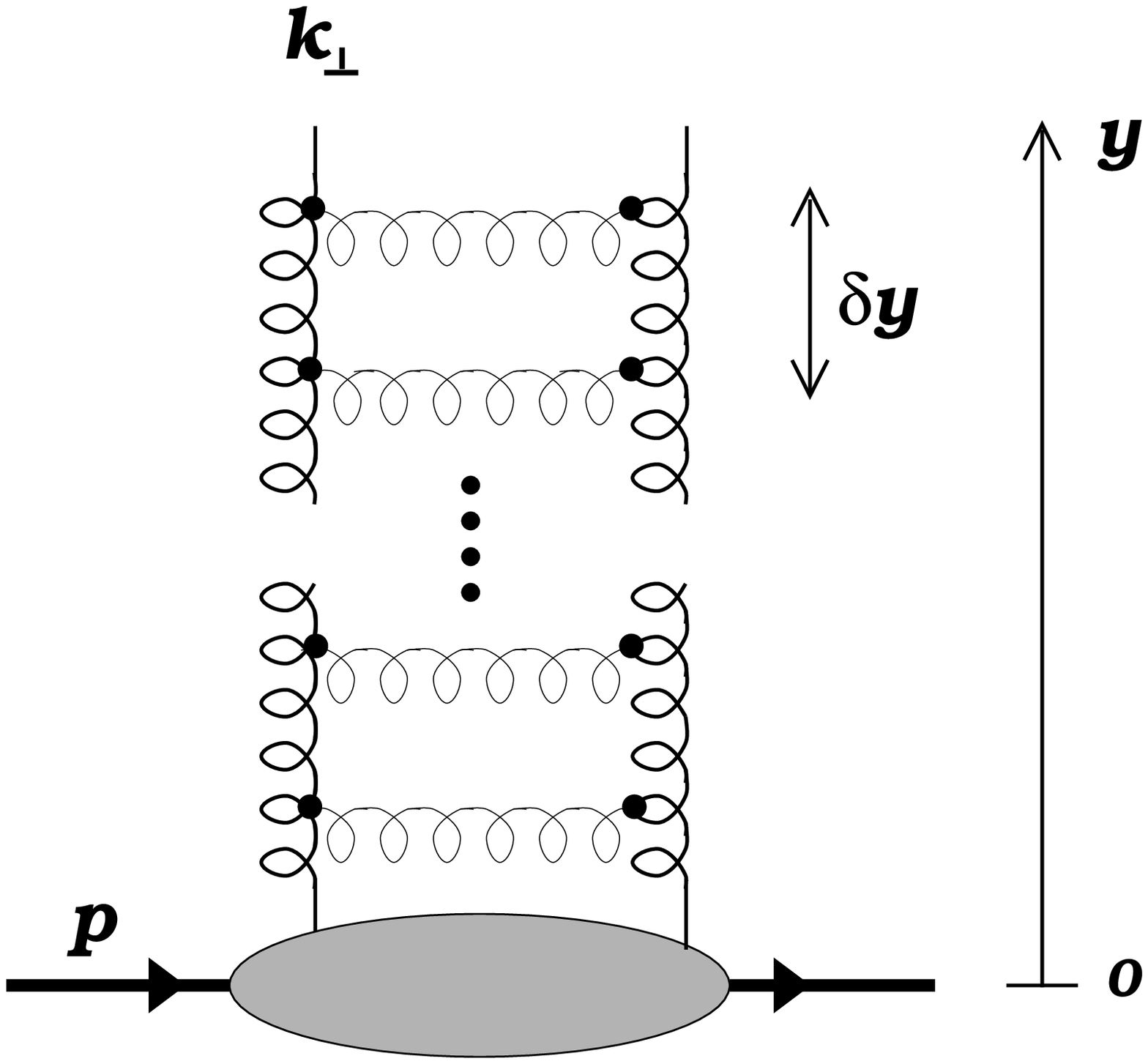}}
\nopagebreak
\begin{center}
\vspace*{1cm}
{\LARGE {\bf Figure 1} }
\end{center} \vspace*{0.5cm}
\vspace*{1cm}
\centerline{\epsfxsize=6cm {\epsfysize=8cm \epsffile{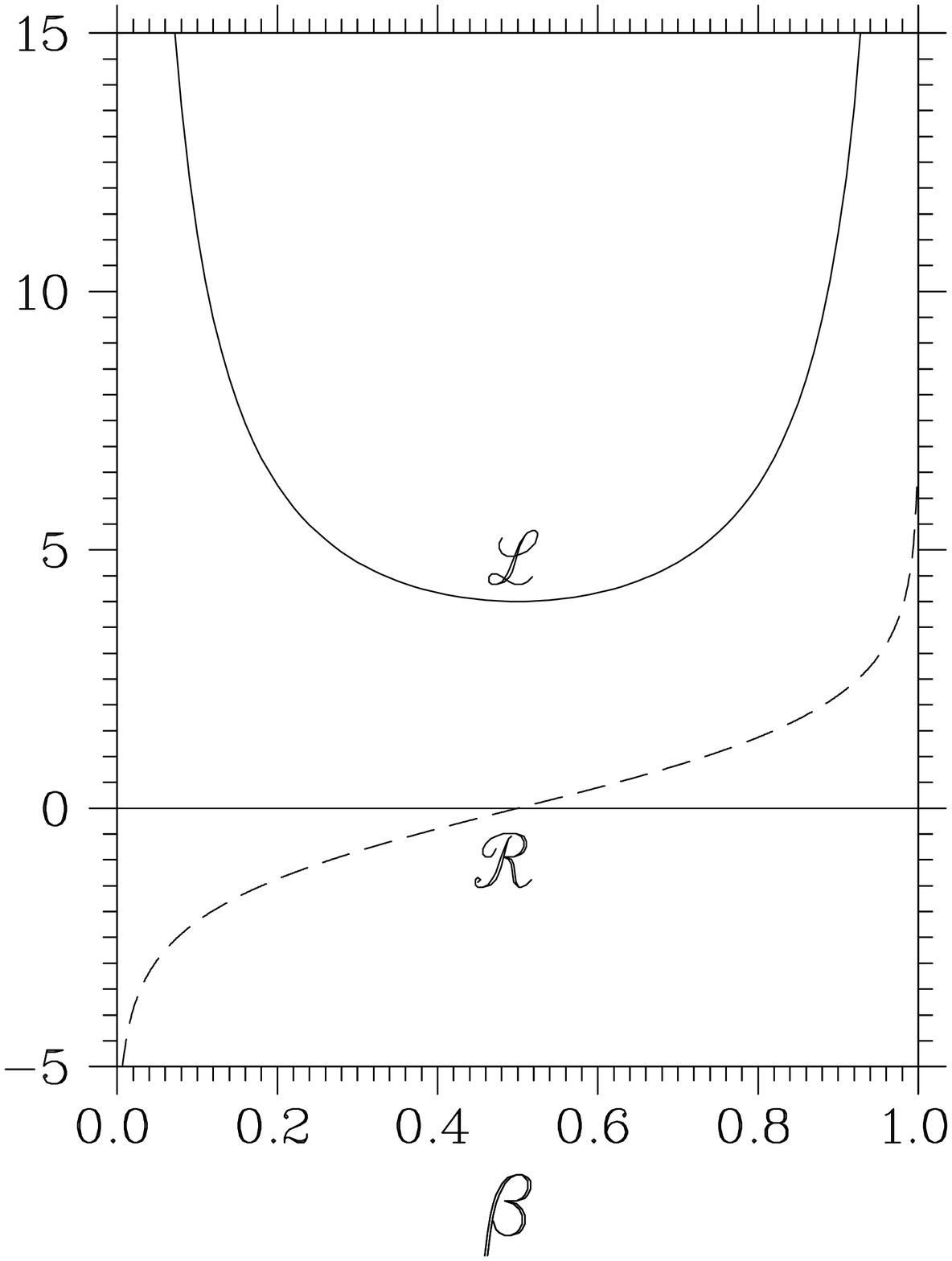}}}
\nopagebreak
\begin{center}
{\LARGE {\bf Figure 2} }
\end{center} \vspace*{0.5cm}
\newpage
\vspace*{1cm}
\begin{tabular}[c]{cc}
{\epsfxsize=6cm {\epsfysize=6cm \epsffile{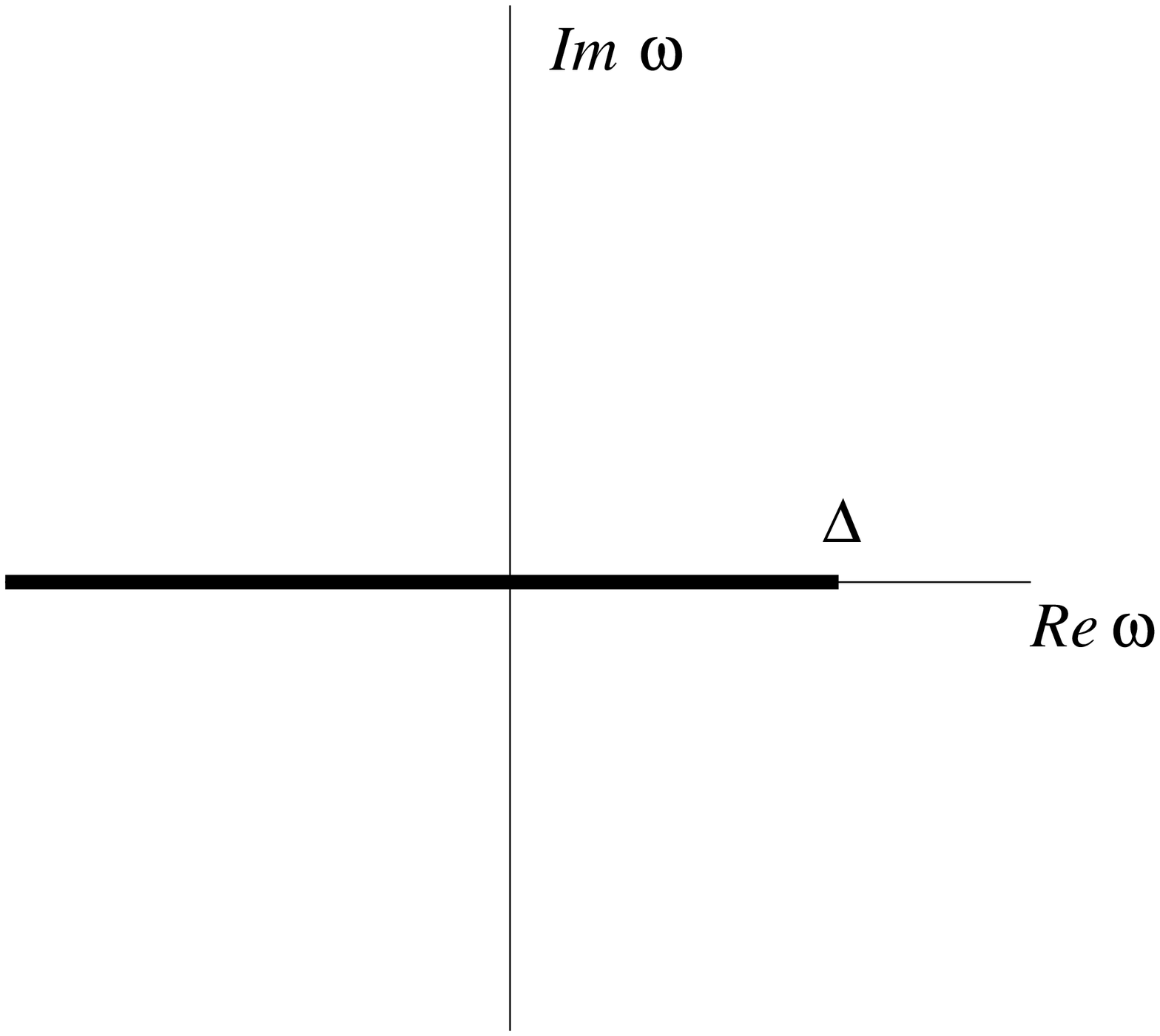}}} &
\hspace{2cm}
{\epsfxsize=6cm {\epsfysize=6cm  \epsffile{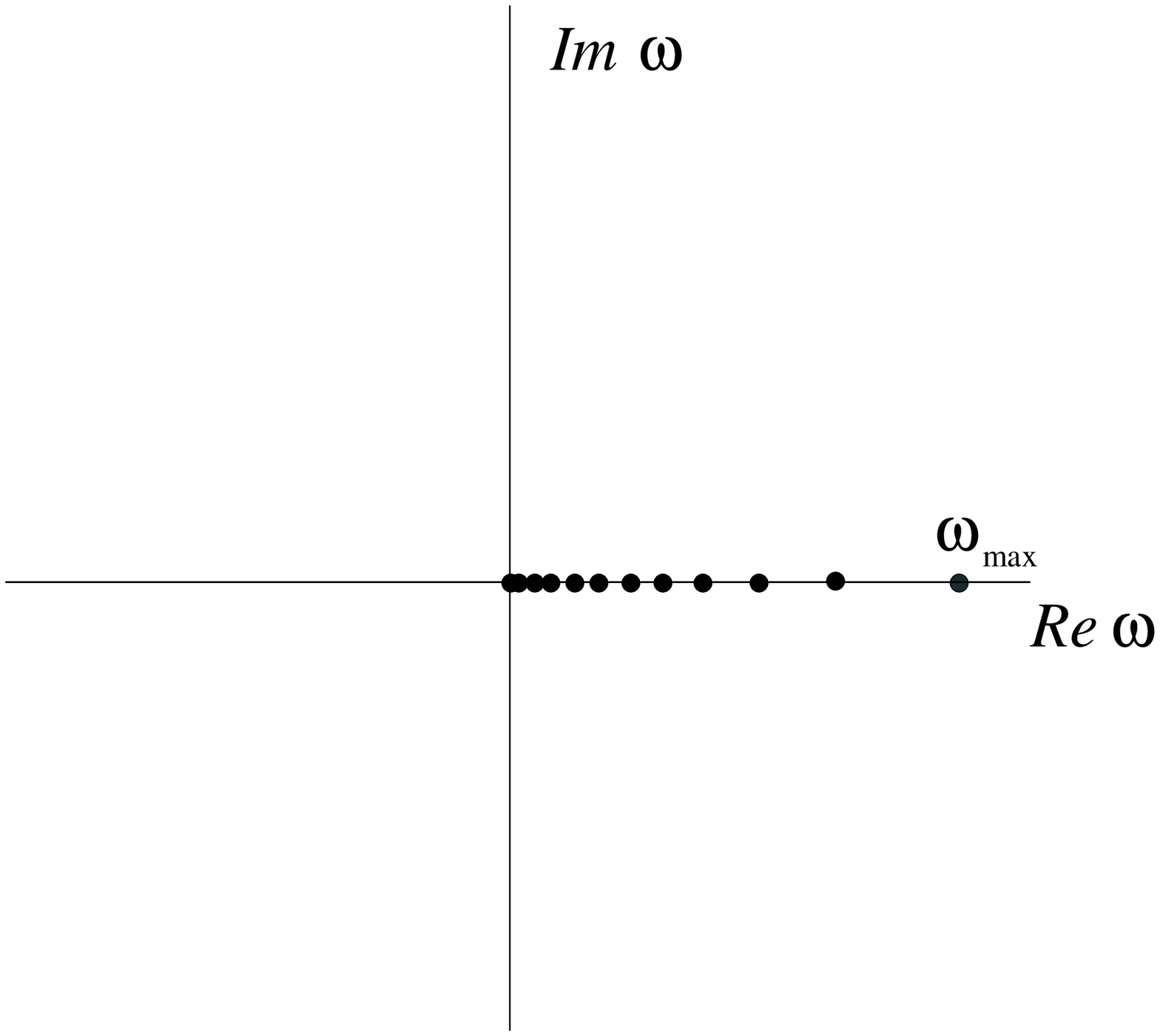}}} \\
{\Large {\bf a.}} & {\Large {\bf b.}}
\end{tabular}
\nopagebreak
\begin{center}
\vspace*{1cm}
{\LARGE {\bf Figure 3} }
\end{center} \vspace*{0.5cm}
\vspace*{1cm}
\centerline{{\epsfxsize=6cm {\epsfysize=6cm  \epsffile{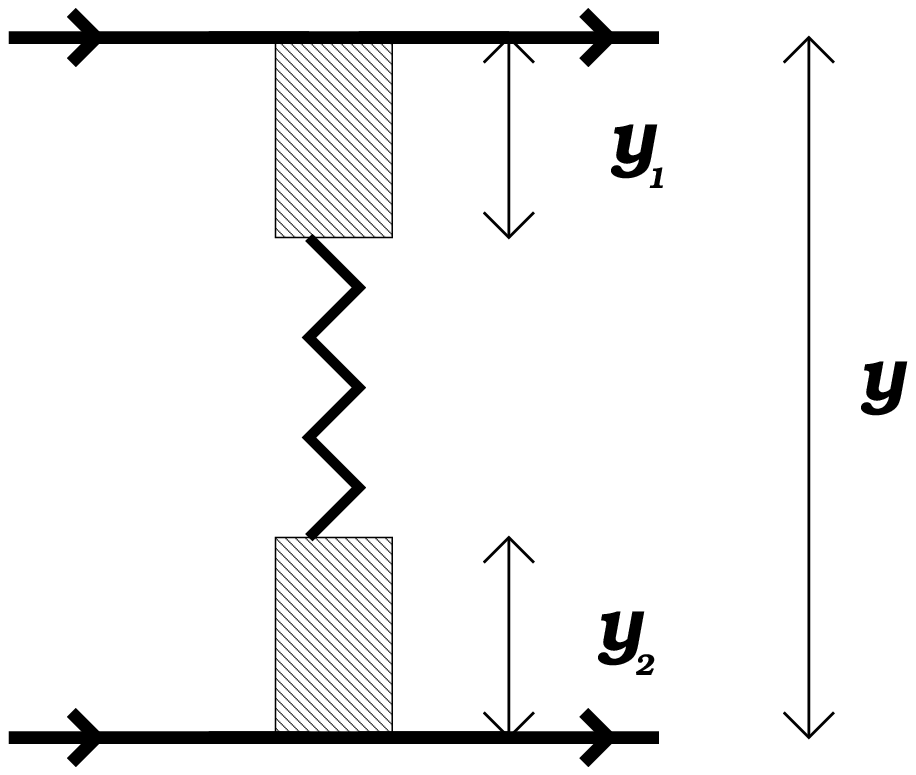}}}}
\nopagebreak
\begin{center}
\vspace*{1cm}
{\LARGE {\bf Figure 4} }
\end{center} 


\begin{references}
\bibitem{BFKL} V.S.Fadin, E.A.Kuraev, L.N.Lipatov,
Sov.Phys.JETP {\bf 44} (1976) 443;
V.S.Fadin, E.A.Kuraev, L.N.Lipatov, Sov.Phys.JETP {\bf 45} (1977) 199;
Y.Y.Balitskii and L.N.Lipatov, Sov.J.Nucl.Phys {\bf 28} (1978) 822
\bibitem{2g_pom} F.F.Low, Phys.Rev {\bf D 12} (1975) 163;
S.Nussinov, Phys.Rev.Lett. {\bf 34} (1975) 1286
\bibitem{next_alpha} V.S.Fadin and L.N. Lipatov, 
Preprint DESY 96-020 (1996)
\bibitem{eff_act} L.N. Lipatov, Nucl.Phys. {\bf B 365} (1991)
614
\bibitem{Lip2} L.N. Lipatov, Sov. Phys. JETP {\bf 63} (1986) 904;
L.N. Lipatov {\em in} { Perturbative Quantum Chromodynamics} Ed.
A.H. Mueller, Advanced Series in High Energy Physics 
(World Scientific Singapore, 1989) 411
\bibitem{Kwe1} J.Kwiecinski, Z.Phys. {\bf C 29} (1985) 561
\bibitem{HanRoss} R.E.Hancock and D.A.Ross, {Nucl.Phys.} 
{\bf B 383} (1992) 575.
\bibitem {NiZa1} N.N.Nikolaev and B.G.Zakharov, Phys.Lett {\bf B 327}
(1994) 157
\bibitem{Levin2} E.M.Levin, TAUP 2221-94, hep-ph 9412345
\bibitem{BrPa} M.A.Braun and C.Pajares, Phys. Lett. {\bf B 287} 
(1992) 154; M.A.Braun and C.Pajares, Nucl.Phys. {\bf B 390} 
(1993) 542
\bibitem{CimCa} G.Camici and M.Ciafoloni, Preprint DFF/260/11/96
(hep-ph/9612235)
\bibitem{jlo} L.P.A. Haakman, O.V. Kancheli, J.H. Koch,
Phys.Lett. {\bf B 391} (1997) 157
\bibitem{erice} O.V. Kancheli, {\em in} Proceedings of the Workshop
``Universality Features in Multihadron Production and the Leading 
Effect''; Erice, Italy (1996)
\bibitem{ColLan1}
J.C.Collins and P.V.Landshoff, Phys.Lett. {\bf B 276} (1992) 196
\bibitem{MFR}  M.F.McDermott, J.R.Forshaw, G.G.Ross,
{Phys.Lett.} {\bf B 349} (1995) 189
\bibitem{Kwe2} J.Kwiecinski, A.D.Martin, P.J.Sutton
Z.Phys. {\bf C 71} (1996) 585
\bibitem{AbraSte} M. Abramowitz and I.A. Stegun, Handbook of 
Mathematical Functions
\bibitem{Lovelace} C.Lovelace Nucl.Phys  {\bf B 95} (1975) 12
\bibitem{KirLip} R.Kirschner and L.N.Lipatov
Z.Phys. {\bf C 45} (1990) 477
\bibitem{GLR} L.V.Gribov, E.M.Levin, M.G.Ryskin, Phys.Rep. {\bf 100} 
(1983) 1; E.M.Levin, M.G.Ryskin, Phys.Rep. {\bf 189} (1990) 267
\bibitem {Mue1} A.H.Mueller, Nucl.Phys. {\bf B 437} (1995) 107
\bibitem{AndGus1} B.Andersson, G.Gustafson, J.Samuelsson,
Nucl.Phys. {\bf B 467} (1996) 443
\bibitem{AndGus2} B.Andersson, G.Gustafson, H.Kharraziha, J.Samuelsson,
Z.Phys. {\bf C 71} (1996) 613
\end{references}
\end{document}